\documentclass[%
 reprint,
 amsmath,amssymb,
 aps,
showkeys,floatfix, pra, aps, amsmath, amssymb, superscriptaddress, twocolumn, longbibliography]{revtex4-2}

\usepackage{graphicx}% Include figure files
\usepackage{dcolumn}% Align table columns on decimal point
\usepackage{bm}% bold math
\usepackage{multirow}
\usepackage{color,outlines}
\usepackage{amsmath}
\usepackage{physics}
\usepackage{anyfontsize} %
\usepackage{gensymb}
\usepackage{colortbl} % For coloring table lines
\usepackage{xcolor} % For color definitions
\usepackage{threeparttable}
\usepackage{graphicx}
\usepackage{ragged2e}
\usepackage{booktabs}
\usepackage{steinmetz}
\graphicspath{{./images/}}
\usepackage{array}   % in preamble
\usepackage{hyperref}% add hypertext capabilities
\begin{document}

\preprint{APS/123-QED}

\title{Modelling and Analysis of Mechanical and Thermal Response of an Ultrastable, Dual-Axis, Cubic Cavity for Terrestrial and Space Applications}% Force line breaks with \\
\author{Himanshu Miriyala}
\thanks{These authors contributed equally to this work.}
\affiliation{Department of Physics, Indian Institute of Technology Tirupati, Yerpedu-517619, Andhra Pradesh, India.}

\author{Rishabh Pal}
\thanks{These authors contributed equally to this work.}
\affiliation{Department of Physics, Indian Institute of Technology Tirupati, Yerpedu-517619, Andhra Pradesh, India.}

\author{Arijit Sharma}
\email{arijit@iittp.ac.in}
\affiliation{Department of Physics, Indian Institute of Technology Tirupati, Yerpedu-517619, Andhra Pradesh, India.}
\affiliation{Center for Atomic, Molecular, and Optical Sciences and Technologies, Indian Institute of Technology Tirupati, Yerpedu-517619, Andhra Pradesh, India.}

% \collaboration{MUSO Collaboration}%\noaffiliation

%  \homepage{http://www.Second.institution.edu/~Charlie.Author}
% \affiliation{
%  Second institution and/or address\\
%  This line break forced% with \\
% }%
% \affiliation{
%  Third institution, the second for Charlie Author
% \affiliation{%
%  Authors' institution and/or address\\
%  This line break forced with \textbackslash\textbackslash
% }%

% \collaboration{CLEO Collaboration}%\noaffiliation

\date{\today}% It is always \today, today,

\begin{abstract}
\vspace{2mm}
\section*{Abstract}

Transportable all-optical atomic clocks represent the next-generation devices for precision time keeping, ushering a new era in encompassing a wide range of PNT (Positioning, Navigation and Timing) applications in the civil and strategic sectors. Their performance relies on ultra-stable, narrow-linewidth lasers, frequency stabilized to a compact portable optical cavity. Among various designs, the cubic spacer-based ultra-stable cavity is particularly well-suited for transportable applications due to its low sensitivity to vibrations, owing to its symmetric geometry and robust mounting structure. While longer cavities offer a lower fundamental thermal noise floor, one needs to strike a balance between transportability and size. In this aspect, the 7.5 cm dual-axis cubic cavity offers a lower fundamental thermal noise floor in comparison to smaller counterparts, while still retaining a reasonable SWaP (Size, Weight and Power) for terrestrial and aerial PNT applications. Its dual-axis design also enables multi-wavelength laser stabilization, making it a promising candidate for future transportable clock applications.
This work presents a detailed study of the 7.5 cm dual-axis cubic cavity using FEM (Finite Element Method) to evaluate its mechanical and thermal stability. We analyze the impact of various geometric factors, mounting forces, and machining imperfections, while also modelling thermal effects such as conduction, radiation, and mirror heating within a vacuum chamber and thermally shielded environment. Our findings provide design insights for developing robust dual-axis optical reference cavities, advancing the deployment of portable atomic clocks for next-generation applications in PNT, geodesy, VLBI (Very Long Baseline Interferometry) and deep space missions.
\end{abstract}


\maketitle

\section{Introduction}\label{sec1}

The pursuit of ever more precise time and frequency standards has transformed modern science and technology, underpinning applications from satellite navigation and telecommunications to tests of the fundamental laws of physics\cite{1,2,3,4,5,6}. Optical atomic clocks have emerged as the most accurate and stable timekeepers, with fractional frequency instabilities reaching the $10^{-18}$ level \cite{7}. Such performance is made possible by interrogating ultranarrow line-width dipole forbidden atomic transitions whose frequencies are stabilized to ultra-stable high-finesse optical resonators. In this context, the frequency stability of the laser and thus of the clock itself is directly determined by the dimension stability of the reference cavity to which it is referenced. The fractional frequency stability of the stabilized laser is directly proportional to the fractional length stability of the cavity. In state-of-the-art laboratory systems, ultra-stable cavities are generally longer operated in vibration-isolated, well temperature-controlled environments, pushing laser instabilities into the low $10^{-16}$ range over 1 s averaging times \cite{Herbers:22}. However, future applications increasingly demand transportable and even space-qualified clock systems, where such environmental isolation is not always feasible. Transportable optical clocks are being developed for inter-laboratory comparisons, precision time transfer in remote locations, spaceborne fundamental physics missions, and navigation systems that extend beyond the limits of present PNT-based systems \cite{pal2024transportable}. In these scenarios, the reference cavity must maintain exceptional stability despite exposure to mechanical shocks, varying orientations, and fluctuating thermal conditions.\\

A key challenge in transportable cavities \cite{pal2024transportable} lies in minimizing acceleration sensitivity without relying solely on external vibration isolation. Among the various approaches to achieving robust frequency references outside controlled laboratory settings, symmetry-based mounting strategies have proven particularly effective in suppressing sensitivity to inertial forces. One of the notable examples is the force insensitive cubic cavity \cite{8,9}, where a cubic spacer is truncated at the vertices and supported at the four vertices in a tetrahedral arrangement. In this geometry, each support force is applied toward the geometric centre of the cube, constraining all translational and rotational degrees of freedom while maintaining force symmetry. This configuration effectively cancels first-order cavity length changes induced by both static preload forces and linear or rotational accelerations. With careful optimization of the vertex truncation depth, such cavities have demonstrated passive acceleration sensitivities as low as 10$^{-11}/g$ \cite{8,10} making them strong candidates for use in transportable and spaceborne optical clock systems \cite{Cole:24}. Building on this foundation, the dual-axis cubic cavity \cite{11,12} extends the concept by incorporating two orthogonal optical axes within the same cubic spacer. This enables simultaneous stabilization of multiple lasers \cite{13} for example, a narrow linewidth clock laser and other cooling or trapping lasers within a single mechanically robust platform.\\

\begin{figure*}[!htb]
\centerline{{\includegraphics[width=0.52\linewidth]{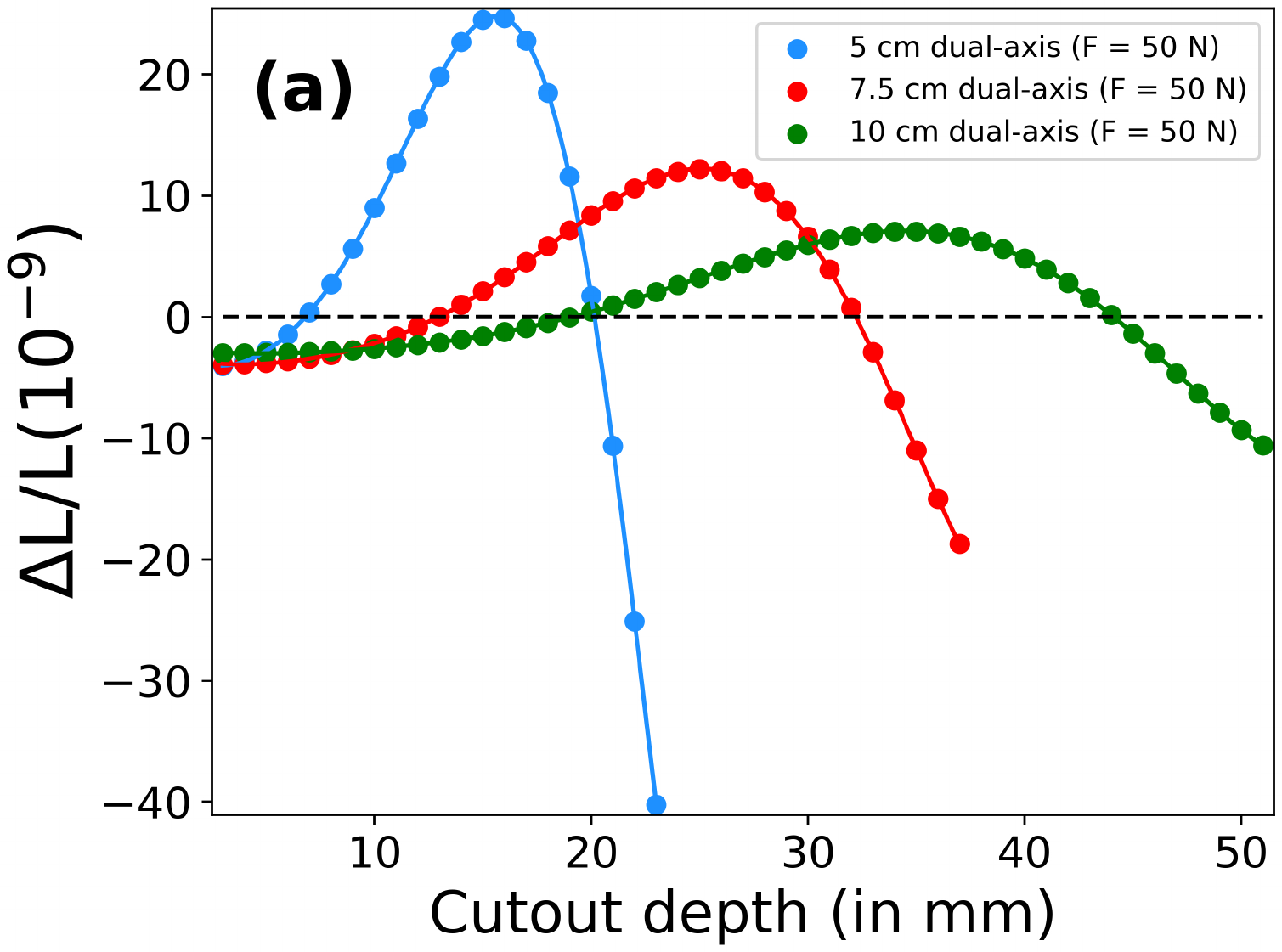}}
\includegraphics[width=0.52\linewidth]{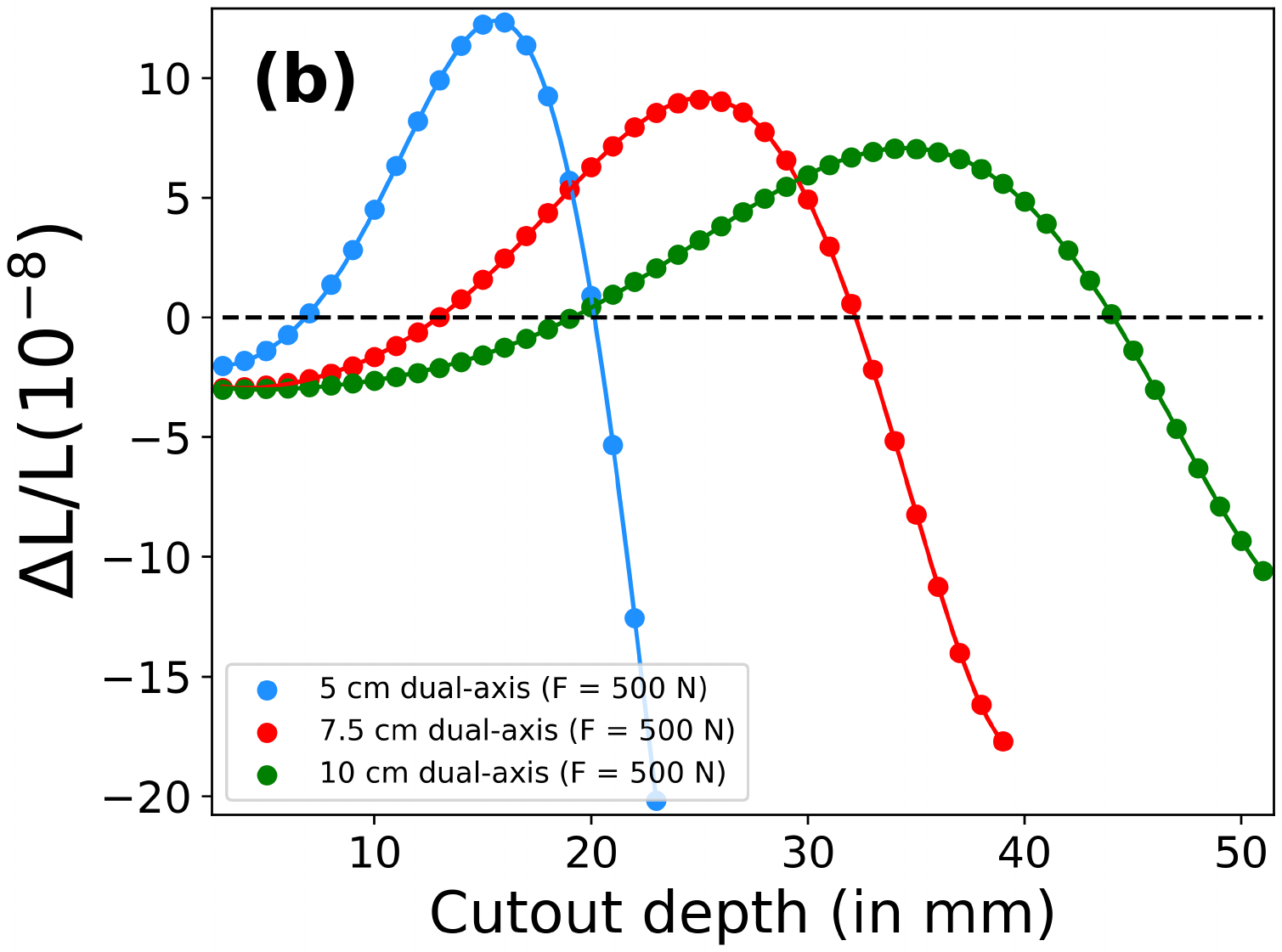}}
\caption{Comparative study of 5 cm, 7.5 cm, and 10 cm dual-axis cubic cavities under supportive forces of (a) 50 N (for terrestrial applications) and (b) 500 N (for space applications).\label{5vs7.5vs10}}
\end{figure*}

Many research groups have demonstrated cubic spacer cavities \cite{8,10,11,12,13,Cole:24}, among which the 5 cm cubic cavity (patented by NPL, UK \cite{8}) has become a prominent choice. The compactness and portability of this 5 cm cubic cavity make it particularly suitable for transportable optical clock applications, positioning it as a leading candidate for future space missions \cite{14,Sanjuan:19,pal2024transportable}. On the other hand, some groups have demonstrated 10 cm cubic cavities \cite{Chen:20,CHEN2023108915,app122412763}, which benefit from a longer cavity length that significantly reduces the thermal noise floor. This makes them a better candidate for transportable systems which require high frequency stability. However, the larger size limits their portability, reducing suitability for transportable or space applications and increasing costs due to greater complexity in thermal management and vibration isolation.\\

Therefore, to strike a balance on size (which determines portability) and noise performance, an analysis comparing different sizes of cubic cavities is essential. We perform a comparative study of 5 cm, 7.5 cm and 10 cm dual-axis cubic cavities, analysing their vibration sensitivity and thermal noise floor. This analysis is conducted for terrestrial applications (with supporting forces of 50 N) and space applications (with supporting forces of 500 N). The sensitivity is determined through the FEA simulation in COMSOL, with the results presented in Figure \ref{5vs7.5vs10}. The thermal noise floor for the three spacer sizes is calculated analytically following the procedure from \cite{14}. Both the simulated sensitivities and the analytical noise floor values are summarized in Table \ref{tab1}.\\

From Table \ref{tab1}, it is evident that both sensitivity and thermal noise performance improve as the size of the cubic spacer (cavity length) increases. Notably, the performance metrics of the 7.5 cm spacer are much closer to those of the 10 cm spacer than to the 5 cm spacer. This demonstrates that the 7.5 cm cubic spacer provides a practical balance: it delivers performance nearly equivalent to the 10 cm option while remaining more cost-effective and compact than 10 cm cavity. Thus, the 7.5 cm design represents an optimal compromise between performance, portability, and cost efficiency.\\

\begin{table*}[!t]
\centering
\renewcommand{\arraystretch}{1.3}  % adjust row height for spacing
\setlength{\tabcolsep}{5pt}        % horizontal padding

\begin{tabular}{|c|c|c|c|c|}
\hline
\textbf{Cubic spacer size} &
\begin{tabular}[c]{c}\textbf{1st zero-crossing} \\ \textbf{cutout}\end{tabular} &
\begin{tabular}[c]{c}\textbf{Sensitivity$^{a}$} \\ \textbf{(for $\boldsymbol{F}$ = 50 N)}\end{tabular} &
\begin{tabular}[c]{c}\textbf{Sensitivity$^{a}$} \\ \textbf{(for $\boldsymbol{F}$ = 500 N)}\end{tabular} &
\textbf{Thermal noise floor} \\
\hline

5 cm   & 6.81 mm  & $2.00\times10^{-9}$ & $9.97\times10^{-9}$ & $7.773\times10^{-16}$ \\
7.5 cm & 12.95 mm & $9.05\times10^{-10}$ & $6.73\times10^{-9}$ & $1.673\times10^{-16}$ \\
10 cm  & 19.13 mm & $4.60\times10^{-10}$ & $4.60\times10^{-9}$ & $9.686\times10^{-17}$ \\

\hline

\end{tabular}

\caption{Comparative analysis of cubic cavity size, sensitivity, and thermal noise floor.}

\vspace{2mm}

{\footnotesize 
$^{a}$Sensitivity increases as the magnitude of supporting forces ($F$) increases. For terrestrial applications, $F=50$ N was considered, while for space applications, $F=500$ N was used.
} 

\label{tab1}
\end{table*}

To ensure the cavity maintains exceptional dimensional and frequency stability under practical conditions, a comprehensive investigation of its design and strategies to mitigate various instabilities is essential. In this work, we present a detailed investigation of the design of the cubic cavity geometry and thermal shield system using finite element analysis (FEA) of a 7.5 cm dual-axis cubic cavity tailored for transportable optical clock applications. We evaluate the influence of individual geometric parameters and also quantify the impact of potential machining errors on the length stability. Numerical thermal modelling is employed to estimate the thermal time constants associated with various heat transfer mechanisms, providing insight into how thermal shielding reduces sensitivity to external temperature fluctuations. Furthermore, we analyse the impact of thermal expansion on the cavity’s length stability and demonstrate how the incorporation of annuli rings can be used to tune the cavity’s minimum expansion temperature towards room temperature.\\

The structure of this article is as follows: Section \ref{sec2} details the configuration of the simulation model developed for this work. Section \ref{sec3} examines the influence of key geometric parameters of the cubic cavity on its fractional length instability ($\Delta L/L$). Section \ref{sec4} investigates the sensitivity of cavity performance to machining tolerances, focusing on two critical parameters: the vertex cutout depth and the bore radius. Section \ref{sec5} introduces a numerical thermal model used to evaluate the thermal time constant of the cavity when enclosed within three thermal shields and a vacuum chamber, accounting for both thermal conduction and localized mirror heating arising from absorption in the high-finesse cavity. Finally, a finite-element analysis (FEA) is performed to determine the effective thermal time constant of the complete system, incorporating the combined effects of conduction, radiation, and mirror heating.\\

Overall, this work contributes to the design of a portable ultrastable cavity that achieves low acceleration sensitivity, mechanical robustness and effective thermal management. These attributes are essential not only for developing transportable cavities used in optical clocks but also for critical high-precision applications such as satellite-based timekeeping, geodesy \cite{4}, and deep-space navigation in demanding and dynamic operational environments.\\

% This work aids greatly to design a transportvble ultrastble cavity which shows ultra-low acceleration sensitivity, thermal stability, and mechanical robustness, providing a compact and versatile frequency reference platform. Such a system is not only well-suited for transportable optical clocks, but also directly applicable to satellite-based timekeeping, geodesy \cite{4} and deep-space navigation where uncompromising frequency stability must be maintained in challenging and dynamic environments.

%Why 7.5 cm dual-axis?Why is length stability important? FEA, Purpose of the simulations. In this work, we conduct a comprehensive investigation of a 7.5 cm dual-axis cubic cavity, using Finite Element Analysis (FEA) simulations. Our primary focus is to examine the influence of various mechanical and thermal factors on the cavity's length stability and overall performance. The paper is organized as follows: in Section \ref{sec2}, we describe the setup of the simulation model. In Section \ref{sec3}, we analyze how various geometric parameters of the cubic cavity affect its fractional length instability($\Delta L/L$). In Section \ref{sec4}, we explore the impact of machining errors in two critical parameters, specifically the cutout depth and bore radius on the fractional length instability.%
 
\section{Setting up the simulation model}\label{sec2}

\begin{figure}[!h]
\centering{\includegraphics[width=0.9\linewidth]{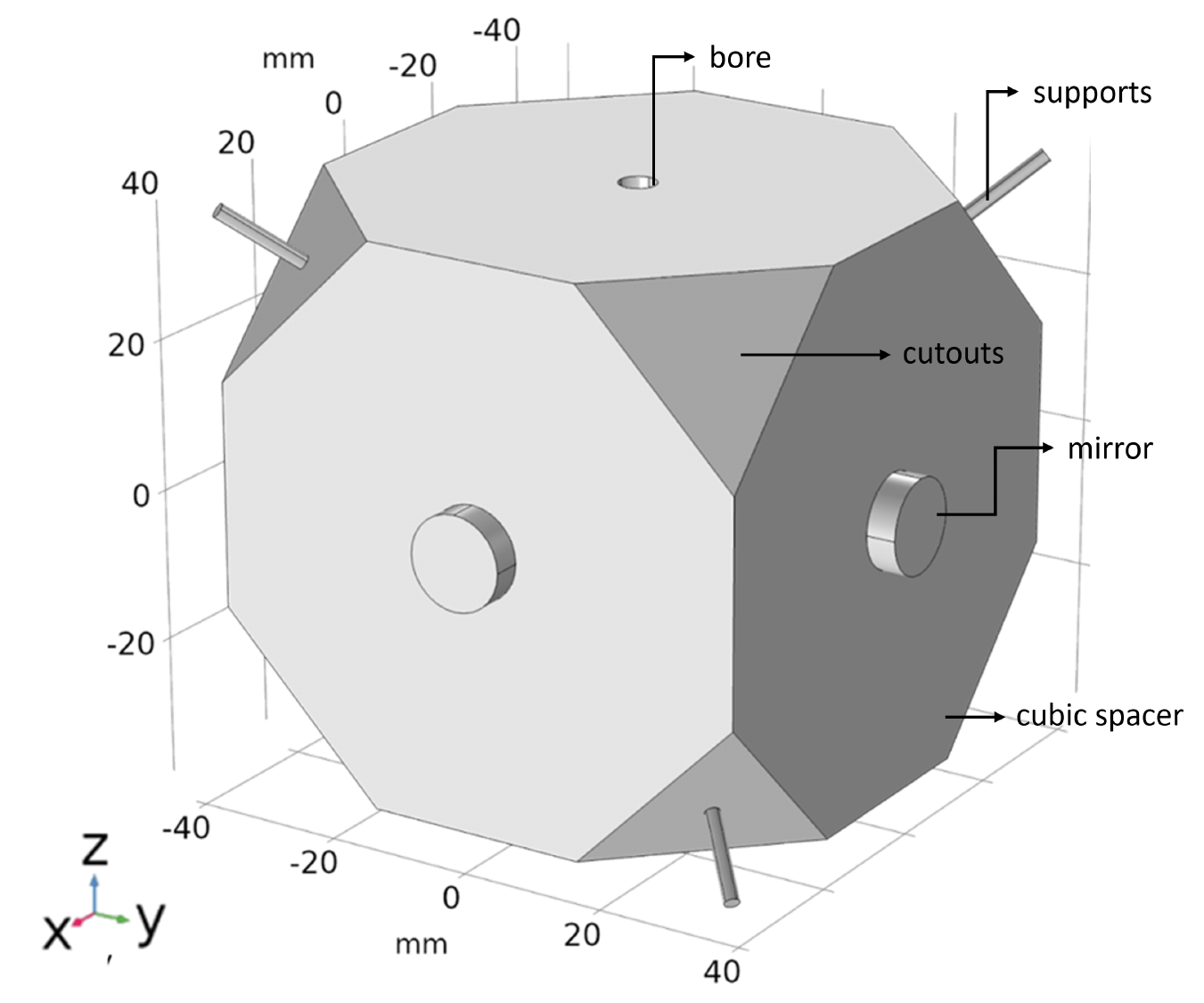}}
\caption{7.5 cm dual-axis cubic cavity design in COMSOL.\label{fig1}}
\end{figure}

In this section, we describe the initial setting for the finite-element analysis (FEA) simulations. We considered the same cavity design as proposed by Webster and Gill \cite{8}. We performed the simulations using commercially available COMSOL software (5.4 version), the model of 7.5 cm dual-axis cubic cavity is as shown in Figure \ref{fig1}. Four mirrors are contacted to the cubic spacer, forming two cavities along two orthogonal axes X and Y. While the Z-axis bore is left open for pumping vacuum. All eight corners of the cube are truncated by a certain depth towards the centre, called the cutout depth. Cylindrical rods were considered to model the four tetrahedral supports. The contact surface of these rods with the spacer is constrained to move only in the longitudinal direction (along the axis of the cylindrical rods). Then, supporting forces are applied by these rods only in the longitudinal direction towards the center of the cube. The geometrical parameters taken for the simulations are tabulated in Table \ref{tab2}.\\

For modelling the ultra-stable cubic cavity, we have selected commonly used materials known for their low thermal expansion and mechanical stability. The spacer material is modelled as Ultra-Low Expansion (ULE) glass, whereas the mirrors as Fused Silica (FS). Cylindrical support rods are modelled as structural steel. The ULE and Fused Silica materials were defined manually by specifying their relevant physical properties, whereas structural steel was taken directly from the COMSOL material library. The key material parameters used in the simulation are summarized in Table \ref{tab3}.

\begin{table}[!t]
\centering
% Increase vertical space between rows
\renewcommand{\arraystretch}{1.3}  
% Increase horizontal padding inside cells
\setlength{\tabcolsep}{8pt}  

\begin{tabular}{|c|c|}
\hline
\textbf{Parameters} & \textbf{Values} \\
\hline
Cube edge length & 75 mm \\
Bore radius ($b$) & 2.55 mm \\
Support radius ($s_r$) & 1 mm \\
Mirror radius ($m_r$) & 6.35 mm \\
Mirror thickness ($m_t$) & 4 mm \\
Cutout depth ($d$) & \textit{free parameter} \\
\hline
\end{tabular}

\caption{Geometric parameters of the cubic cavity used in the simulation.}
\label{tab2}
\end{table}

\begin{table}[!h]
\centering
% Increase vertical space between rows
\renewcommand{\arraystretch}{1.3}  
% Increase horizontal padding inside cells
\setlength{\tabcolsep}{8pt}  

\begin{tabular}{|c|c|c|}
\hline
\textbf{Parameters} & \textbf{ULE} & \textbf{Fused Silica} \\
\hline
Density (kg/m$^3$) & 2203 & 2210 \\
Young's Modulus (GPa) & 67.7 & 73.1 \\
Poisson ratio & 0.17 & 0.17 \\
\hline
\end{tabular}

\caption{Material parameters used in the simulation.}
\label{tab3}
\end{table}

\begin{figure*}[htb]
\centerline{\includegraphics[width=0.49\linewidth]{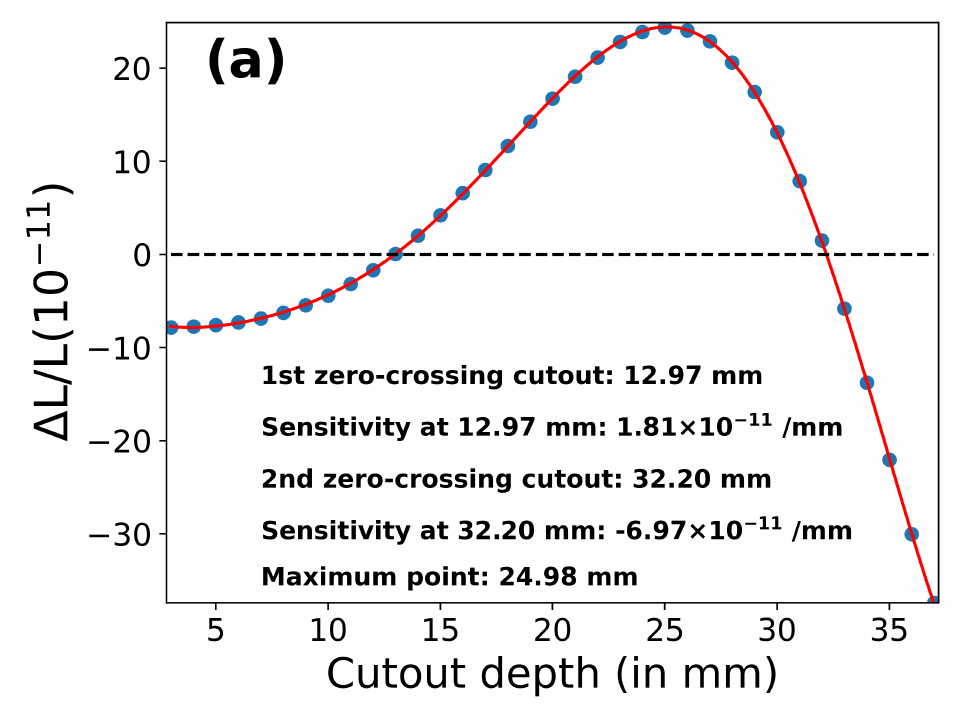}
 \includegraphics[width=0.49\linewidth]{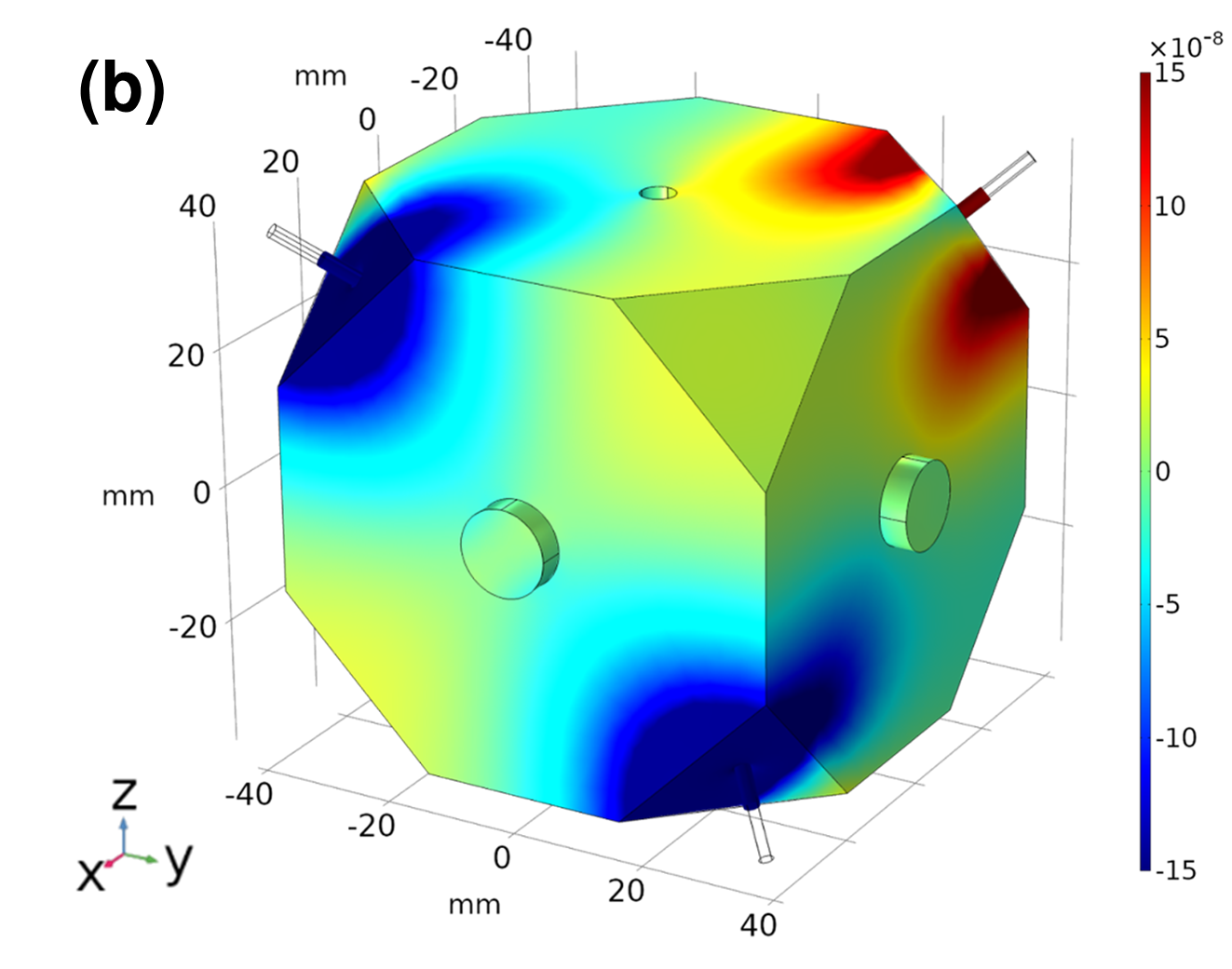}}
\caption{(a) Effect of cutout depth on the length stability ($\Delta L/L$). A 6th-order polynomial fit is applied to the data to determine the zero-crossing cutout depths and corresponding sensitivities. (b) Magnitude of axial displacement (in mm) along the x-direction due to deformation for the first zero-crossing cutout 12.97 mm when supporting forces are applied.\label{fig2}}
\end{figure*}

Under the influence of supporting forces, the cubic cavity experiences deformation, resulting in fluctuations in the cavity length $L$ (distance between the mirrors). However, as described by Webster and Gill \cite{8}, this cavity length instability induced by these supporting forces can be nullified by optimizing the depth of the cutouts. Using COMSOL simulations, we simulated how the relative length fluctuations $\Delta L/L$ change with varying cutout depths, as illustrated in the Figure \ref{fig2}. Notably for our 7.5 cm dual-axis cubic cavity, the length stability ($\Delta L/L$) reaches zero at two specific cutout depths 12.97 mm and 32.20 mm, referred to as zero-crossing cutouts. The slope of the plot at these points indicates the cavity’s sensitivity to machining errors for a particular supporting force exerted by the supports. Of the two zero-crossing cutouts, the first exhibits lower sensitivity than the second, making it the preferred and optimal cutout depth for best cubic cavity performance.\\

\begin{figure*}[htb]
\centerline{\includegraphics[width=0.34\linewidth]{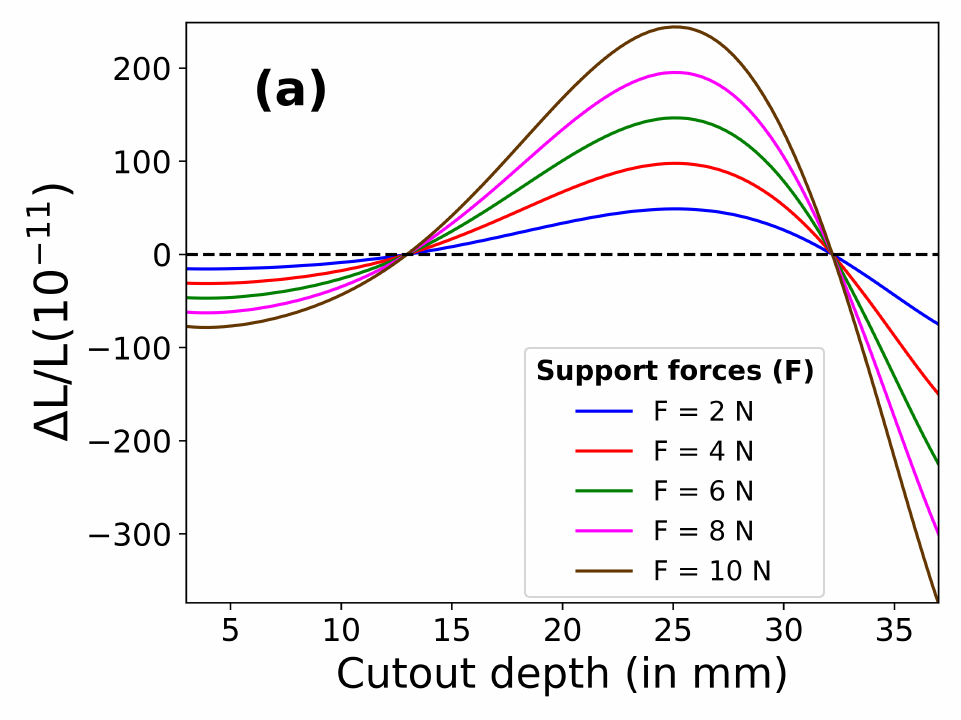}
 \includegraphics[width=0.34\linewidth]{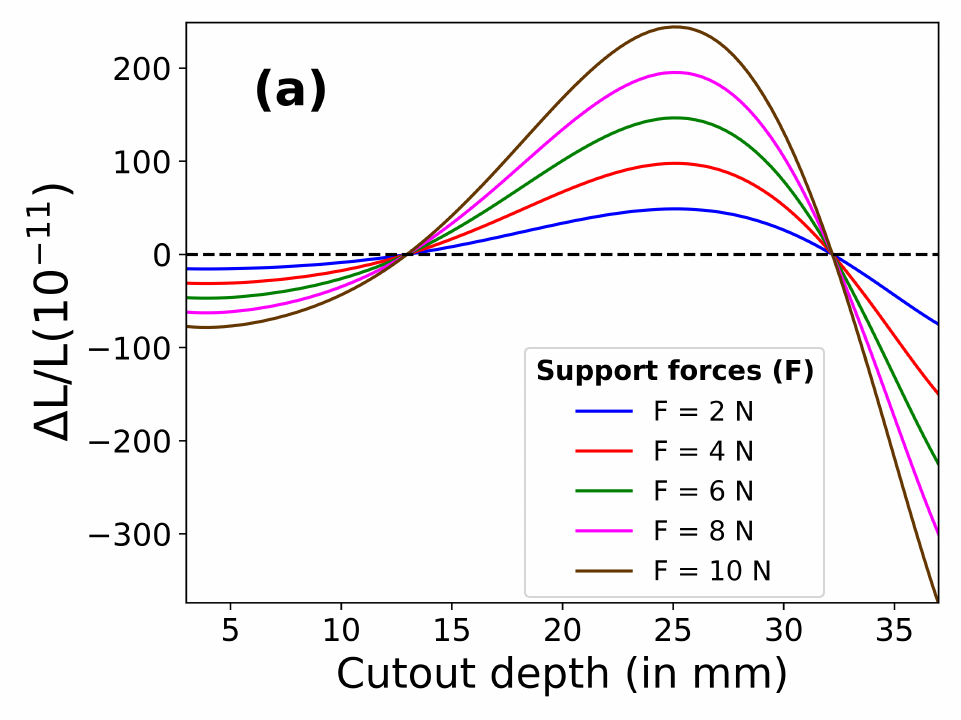}
 \includegraphics[width=0.3\linewidth]{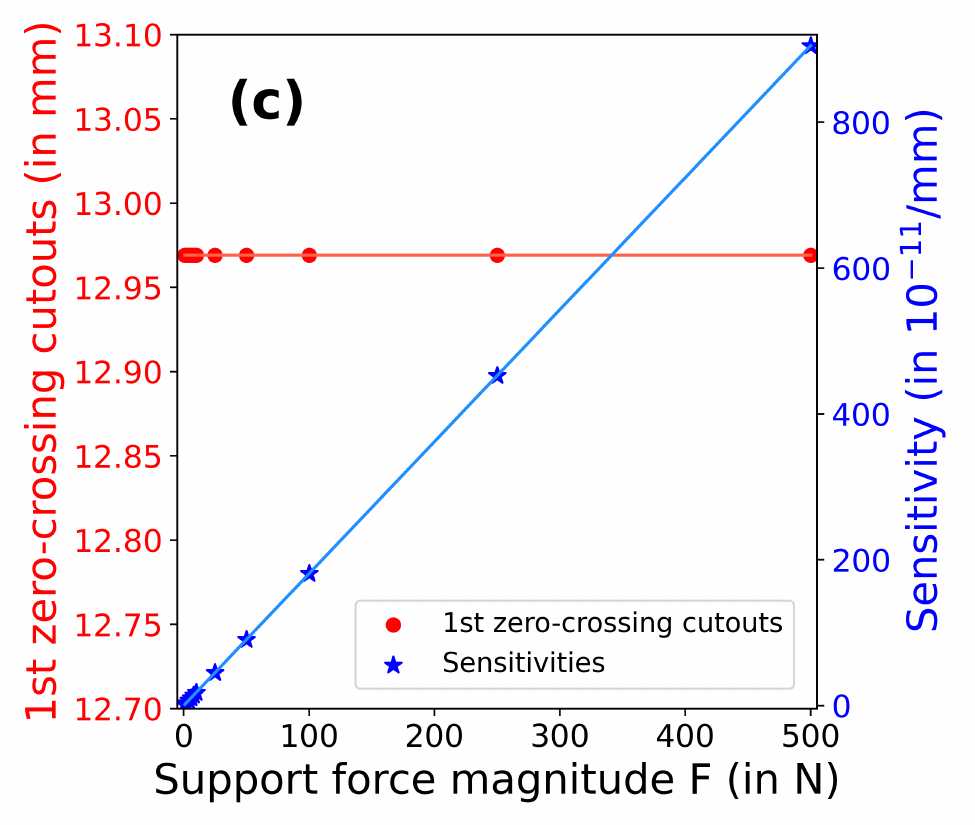}}
\caption{Effect of supporting forces: (a) in the 1-10 N range, (b) in the 25-500 N range, and (c) on zero-crossing cutouts and sensitivity (The lines in this figure are provided for visual guidance only and do not represent the relationship between the data points).\label{fig3}}
\end{figure*}

In the absence of any external acceleration on the cavity, the sensitivity is mainly influenced by the magnitude of supporting forces. To examine the effect of supporting forces applied by the supports, the force magnitude was varied across two ranges: from 2 N to 10 N for lower forces, and from 50 N to 500 N for higher forces. As shown in the Figure \ref{fig3}, the zero-crossing cutouts remain unchanged, confirming their independence from the applied support forces. In contrast, the sensitivity increases linearly with the magnitude of the support forces.\\

In the next section, we discuss the effect of various geometric parameters on the length stability of the cubic cavity as the cutout depth varies and how each of these parameters impacts the first zero-crossing cutout and sensitivity.

\section{Influence of geometric parameters of the cavity}
\label{sec3}

\begin{figure*}[htb]
\centering{\includegraphics[width=0.32\linewidth]{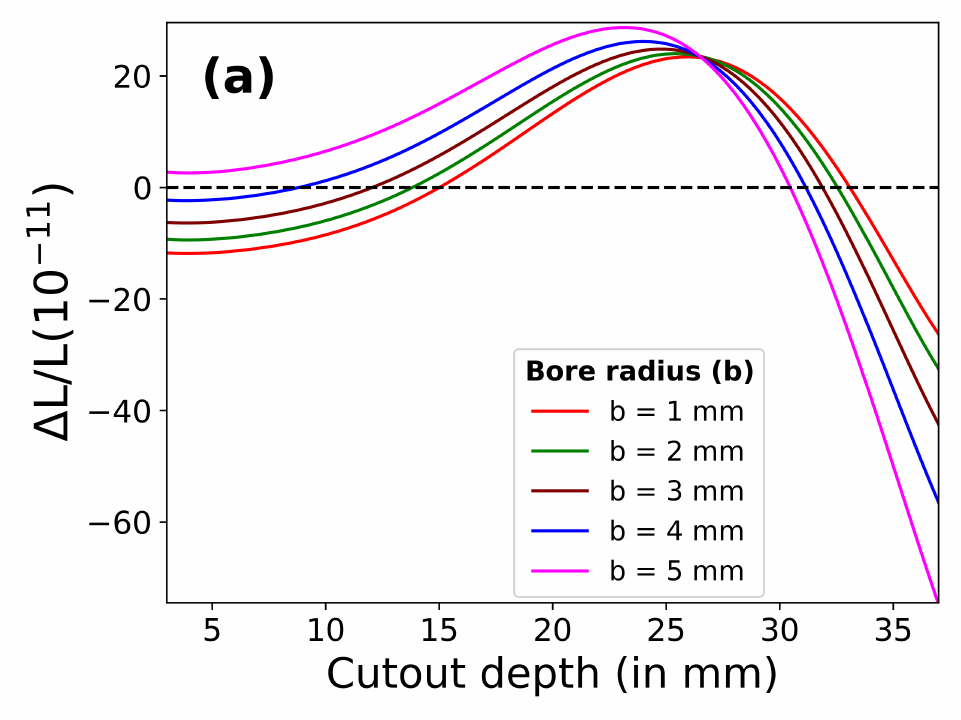}
 \includegraphics[width=0.32\linewidth]{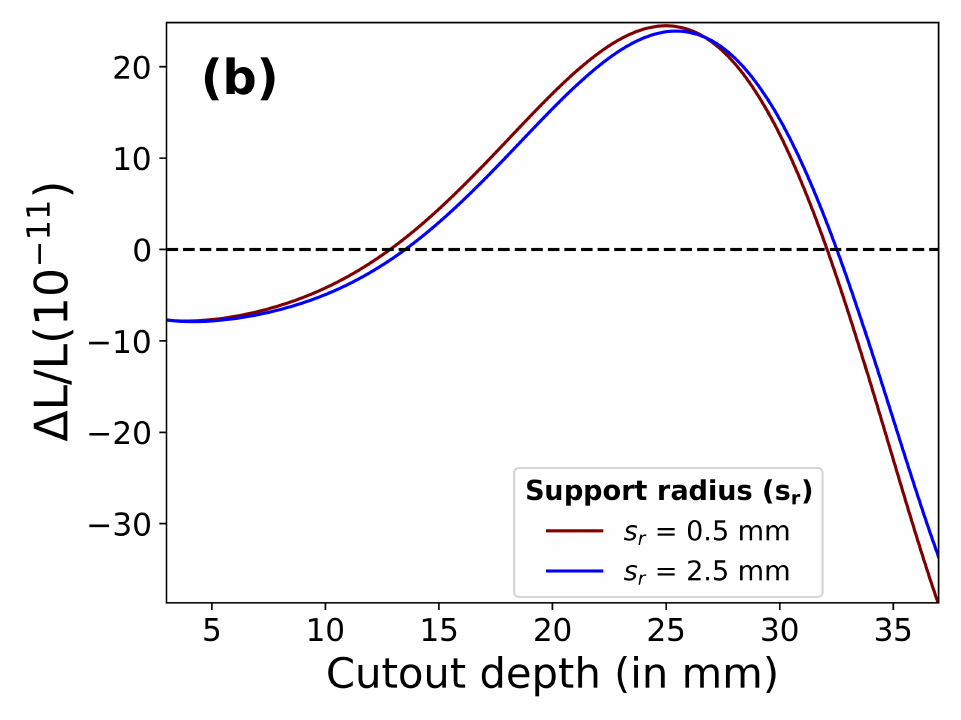}
 \includegraphics[width=0.32\linewidth]{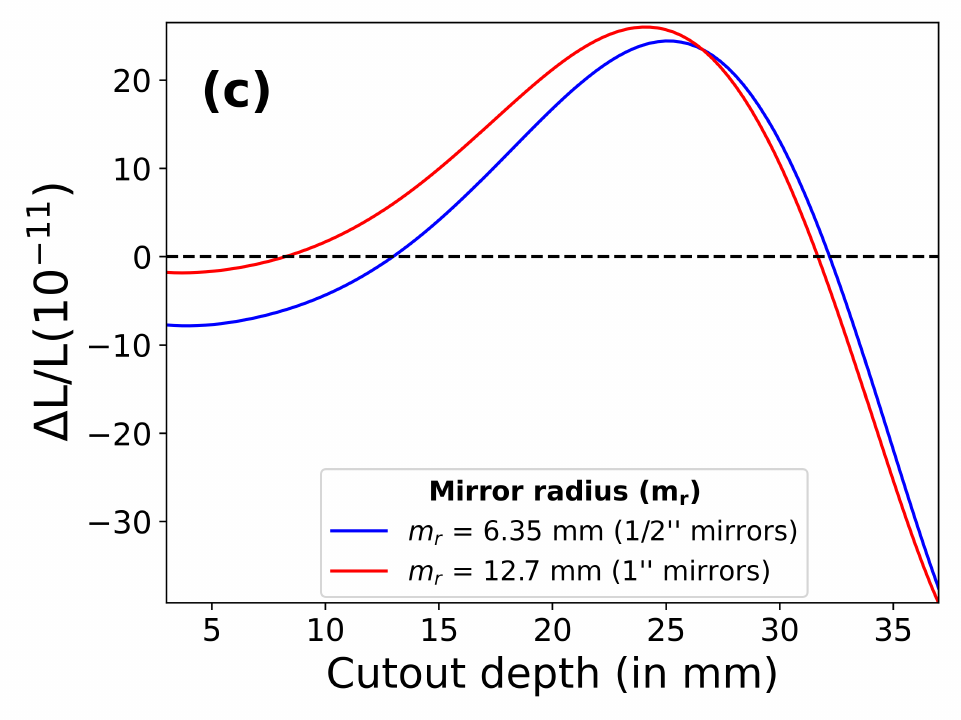}
 \includegraphics[width=0.32\linewidth]{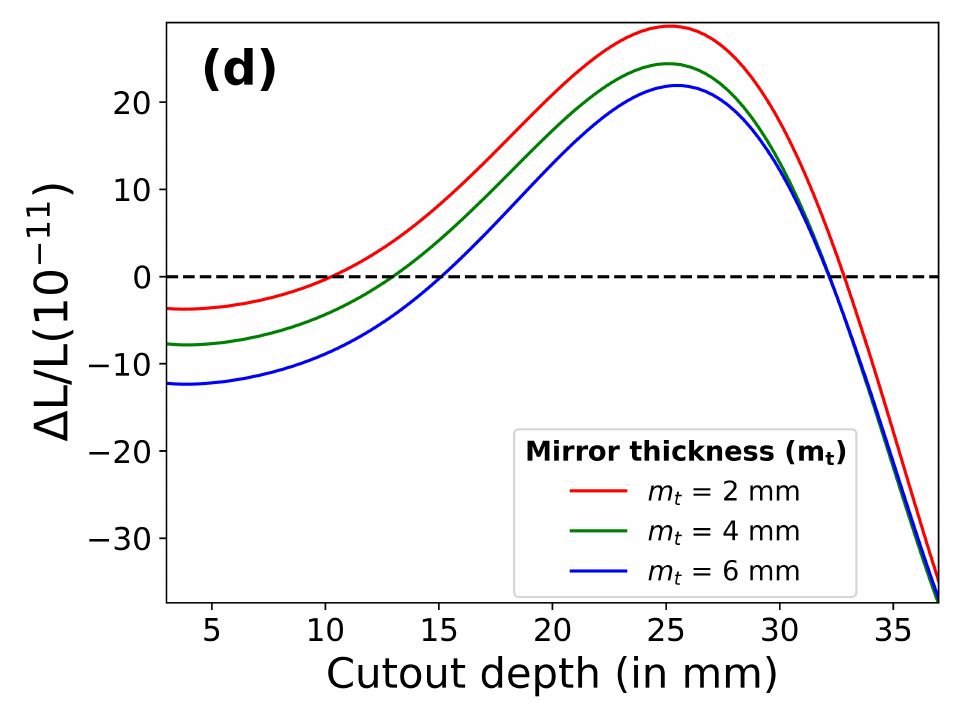}
 \includegraphics[width=0.32\linewidth]{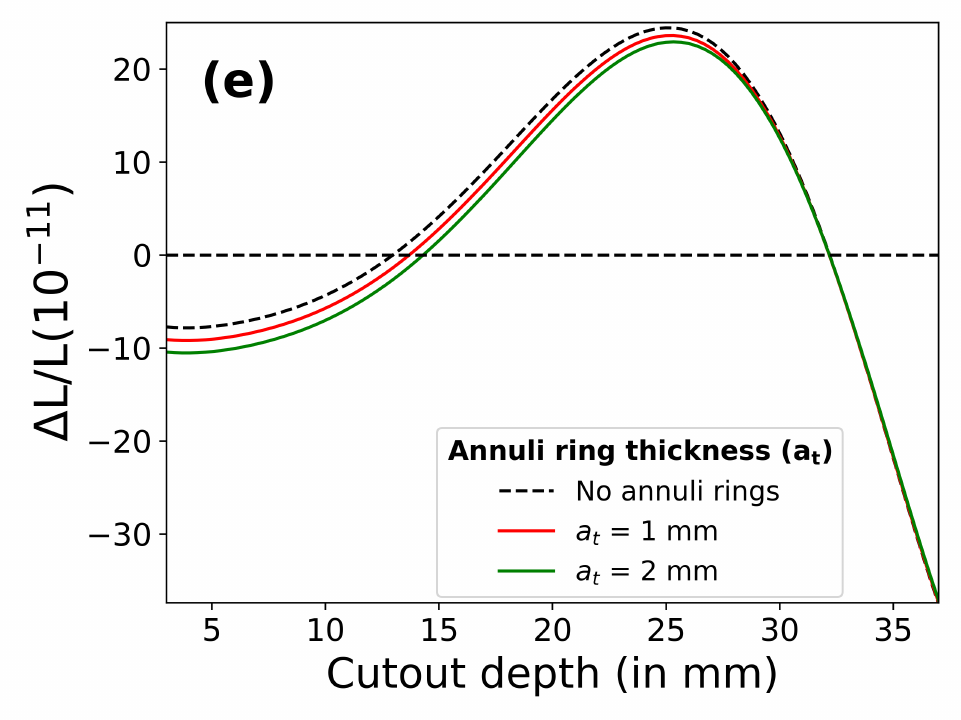}}
 \includegraphics[width=0.32\linewidth]{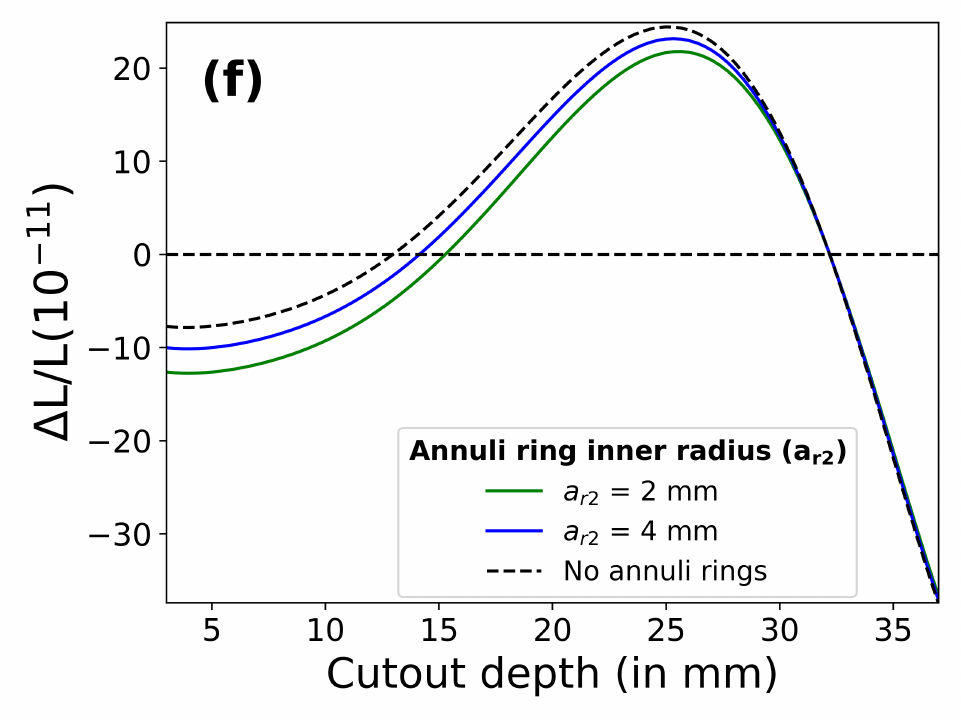}
\caption{Effect of geometric parameters on the length stability of the cubic cavity: (a) Bore radius $b$, (b) Support radius $s_r$, (c) Mirror radius $m_r$, (d) Mirror thickness $m_t$, (e) Annular ring thickness $a_t$, and (f) Annular ring inner radius $a_{r2}$. Each parameter is systematically varied to assess its influence on the cavity’s fractional length change while keeping other parameters constant.\label{fig4}}
\end{figure*}

Figure \ref{fig4} shows how the fractional length instability $\Delta L/L$, varies in response to changes in different geometrical parameters. While studying the influence of annuli ring parameters, annular rings made of ULE material were incorporated by attaching them to the four mirrors in addition to the model shown in Figure \ref{fig1}. Among all these parameters, the most significant influence on fractional length instability comes from the bore radius, while the support radius exhibits the least impact. The effects of these geometrical parameters on the first zero-crossing cutouts and sensitivities are analyzed in detail in the following subsections.
% \subsection{Impact on 1st zero-crossing cutouts and sensitivities}

\subsection{Effect of bore radius}

\begin{figure}[htb]
\centering{\includegraphics[width=0.8\linewidth]{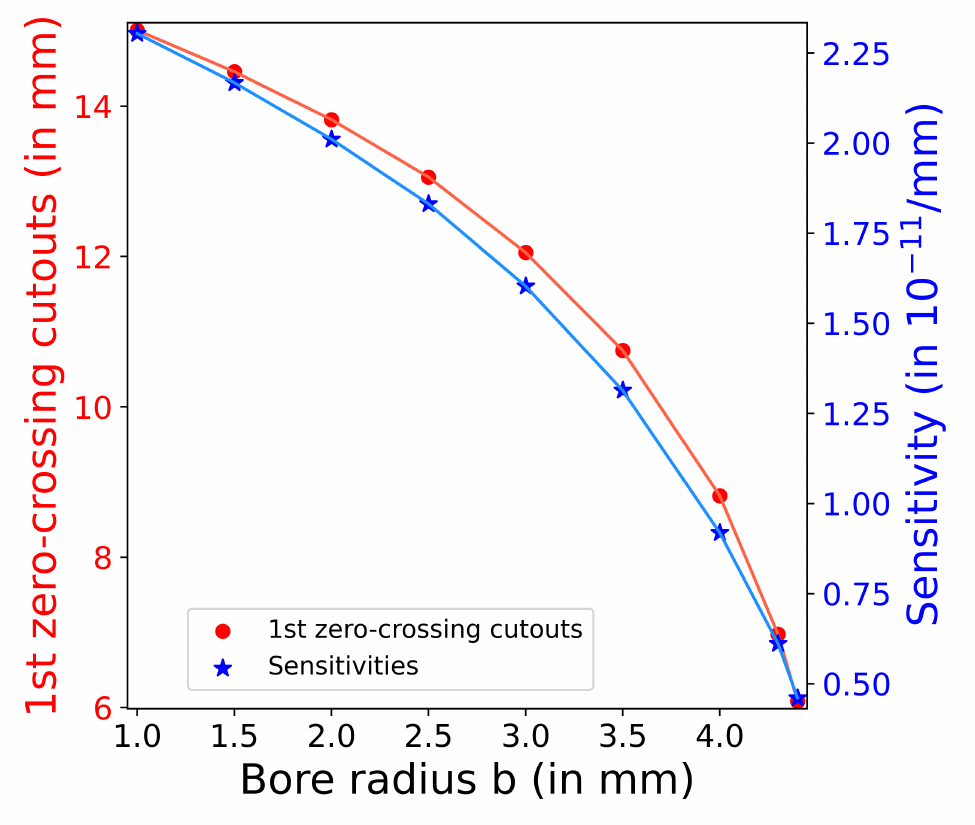}}
\caption{Effect of bore radius ($b$) on first zero-crossing cutouts and sensitivities. The lines in this figure are provided for visual guidance only and do not represent relationship between the data points.\label{fig5}}
\end{figure}

One of the key parameters that strongly influences the fractional length stability is the bore radius ($b$). Figure \ref{fig5} illustrates how the first zero-crossing cutout and sensitivity decrease as the bore radius increases from 1 mm to 6 mm. However it is important to note that, beyond a critical threshold of $b > 4.4$ mm, the first zero-crossing cutout disappears entirely. Since the cavity must be machined precisely to align with the first zero-crossing cutout, it is crucial to select an appropriate bore radius. At the same time, the bore radius should not be too small, as it must accommodate the laser beam's entry. Therefore, selecting a mid-value within the optimal range of 2 mm $< b <$ 3.5 mm is the preferred choice for the bore radius.

\subsection{Effect of support radius}

\begin{figure}[htb]
\centering{\includegraphics[width=0.8\linewidth]{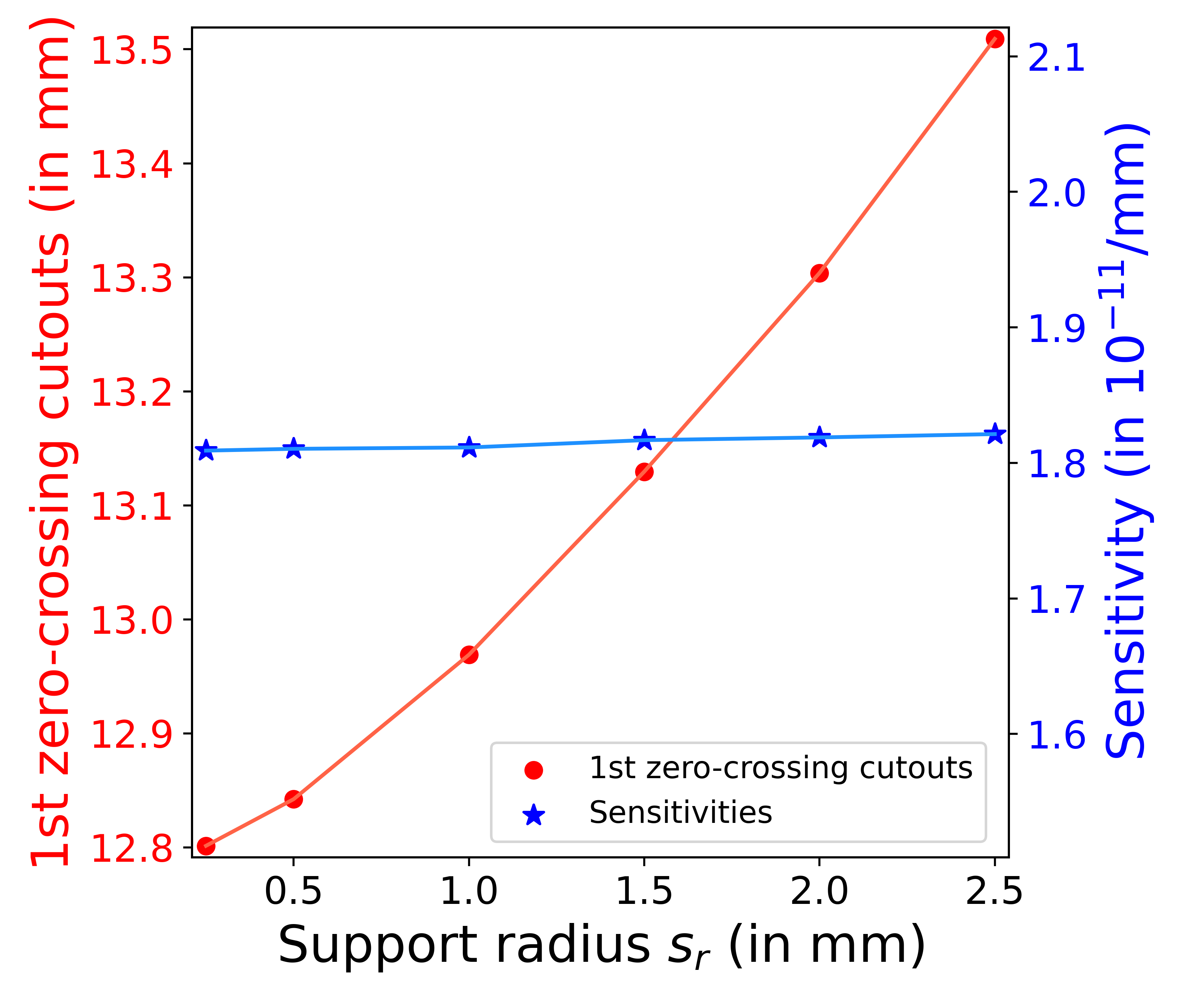}}
\caption{Effect of support radius ($s_r$) on first zero-crossing cutouts and sensitivities. The lines in this figure are provided for visual guidance only and do not represent relationship between the data points.\label{fig7}}
\end{figure}

Figure \ref{fig7} presents the analysis of the first zero-crossing cutout and sensitivity as the supporting rod radius ($s_r$) varies from 0.25 mm to 2.5 mm. The results indicate that while the zero-crossing cutout increases increasing support radius, sensitivity remains nearly constant. Mechanically, this suggests that variations in support radius have minimal impact on sensitivity. From a thermal perspective, however, larger support radii increase thermal conduction from the supports to the cavity spacer, potentially compromising thermal stability. Therefore, having a smaller support radius in the range of $0.25-1$ mm is preferable, as it minimizes thermal conduction. Practically, supporting rods often incorporate Viton balls as intermediates between the rod and the cavity to provide thermal insulation and elastic decoupling from vibrations. Due to their elasticity, the effective contact radius at the cavity interface can vary, affecting the zero-crossing cutout. Roughly, a 0.5 mm change in support radius corresponds to about a 0.1 mm change in zero-crossing cutout. Since smaller support radii cause smaller variations in zero-crossing cutout, maintaining a smaller effective contact radius remains beneficial.

\subsection{Effect of mirror radius}

We evaluated the impact of mirror radius on the length stability ($\Delta L/L$) of a cubic cavity by analyzing mirrors with radii of 6.35 mm (half-inch diameter) and 12.7 mm (full-inch diameter). As shown in Figure \ref{fig4} (c), increasing the mirror radius from half-inch to full-inch reduces the zero-crossing cutout from 12.97 mm to 8.23 mm, and the sensitivity decreases from $1.8\times10^{-11}$/mm to $7.8\times10^{-12}$/mm. 
 
\subsection{Effect of mirror thickness}

\begin{figure}[htb]
\centering{\includegraphics[width=0.8\linewidth]{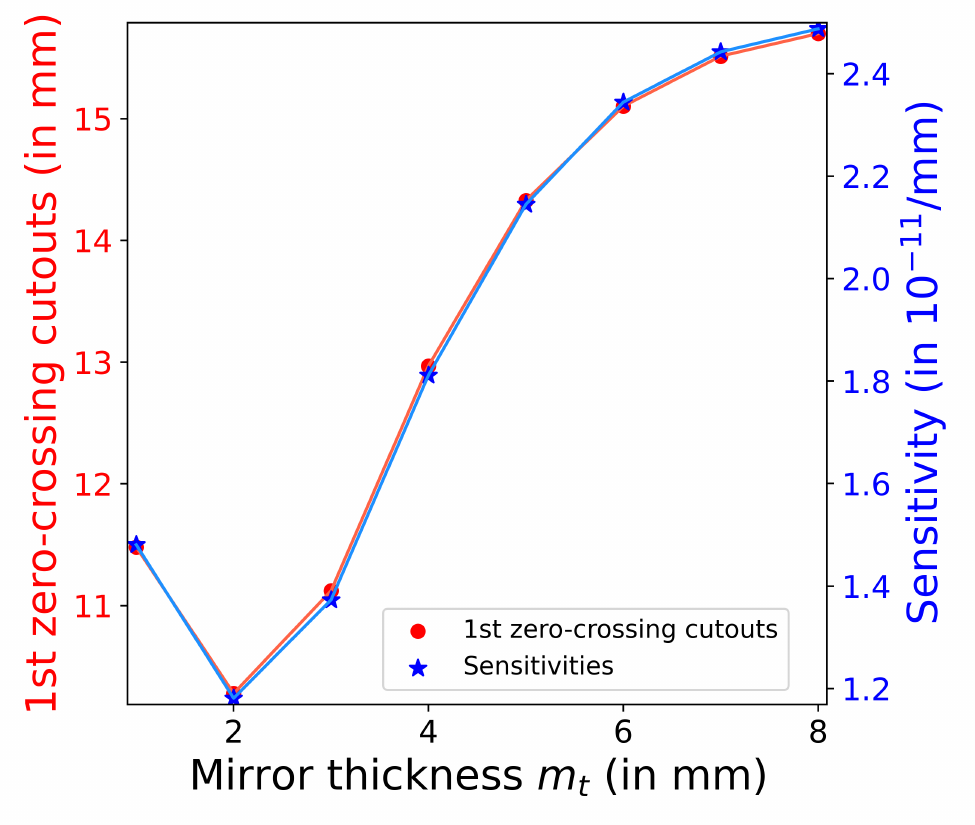}}
\caption{Effect of mirror thickness ($m_t$) on first zero-crossing cutouts and sensitivities. The lines in this figure are provided for visual guidance only and do not represent relationship between the data points.\label{fig6}}
\end{figure}

Figure \ref{fig6} illustrates how the first zero-crossing cutouts and their corresponding sensitivities change as $m_t$ increases from 1 mm to 8 mm. For lower values of $m_t$, both parameters decrease to a minimum at 2 mm, then increase steadily, reaching a maximum at 8 mm.

\subsection{Effect of annuli ring parameters}

Legero et al. \cite{Legero:10} demonstrated that annular rings offer an effective solution for correcting the coefficient of thermal expansion (CTE) mismatch between ultra-low expansion (ULE) and fused silica (FS) mirrors. Since then, this technique has been widely implemented in ultrastable cavities by attaching ULE annular rings to the rear surface of FS mirrors.\\

\begin{figure}[htb]
\centering{\includegraphics[width=0.8\linewidth]{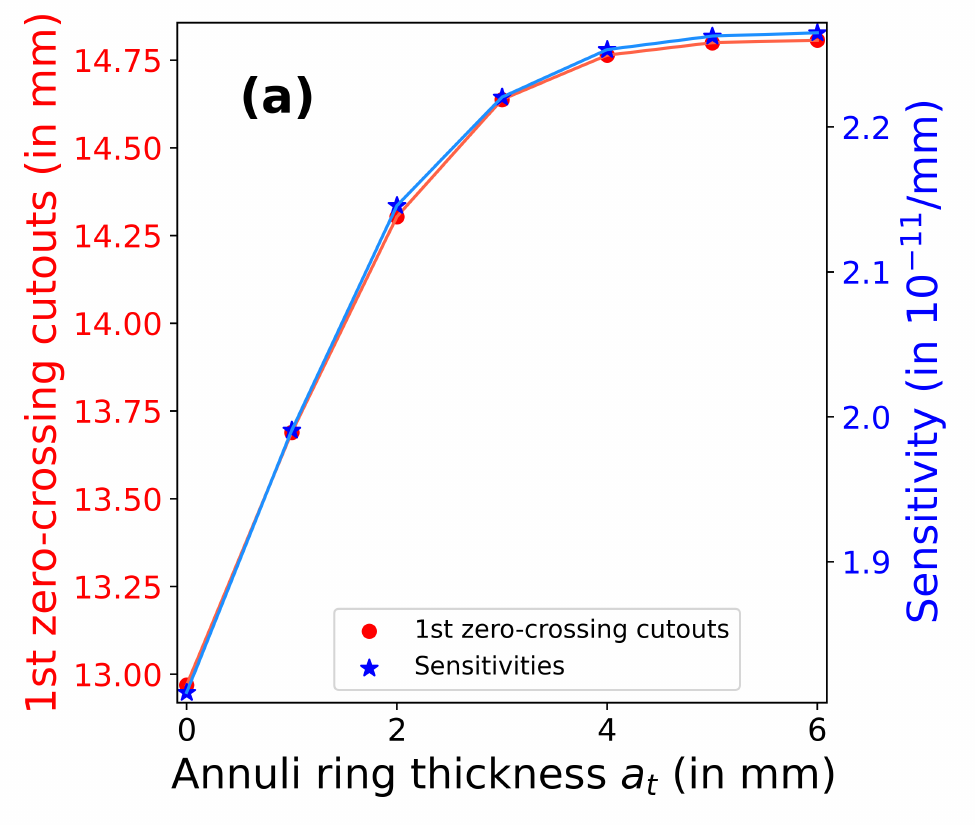}
 \includegraphics[width=0.8\linewidth]{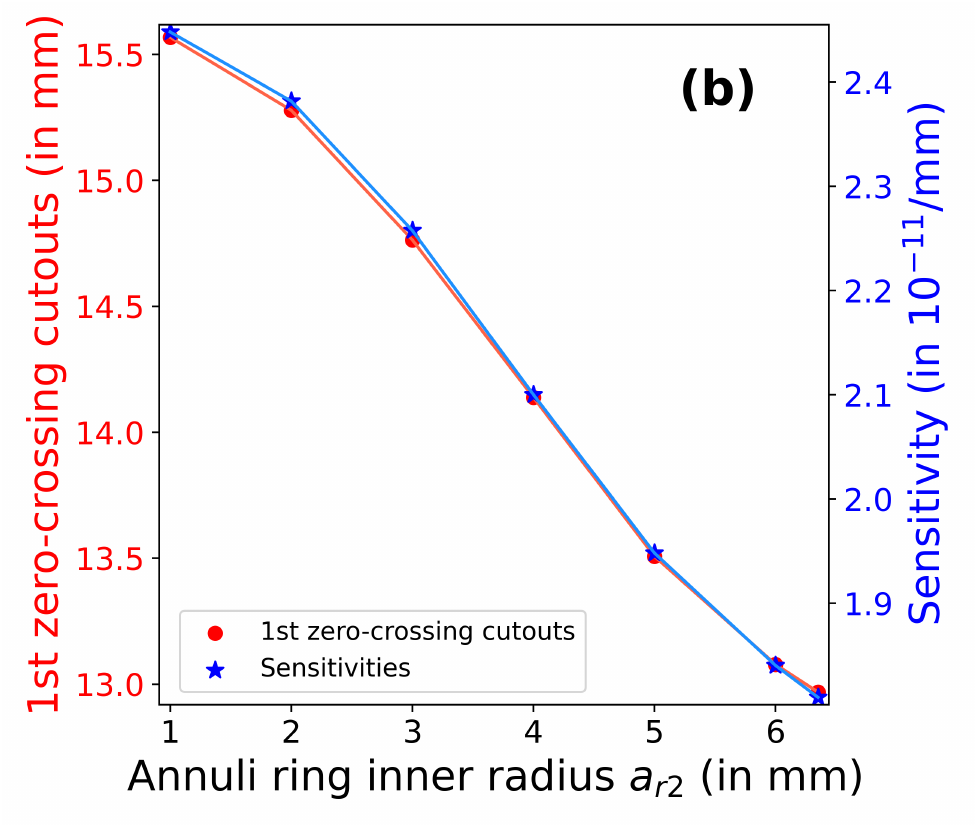}}
\caption{Effect of (a) annuli ring thickness $a_t$ and (b) annuli ring inner radius $a_{r2}$ on first -zero-crossing cutouts and sensitivities. The lines in these figures are provided for visual guidance only and do not represent relationship between the data points.\label{fig8}}
\end{figure}

To examine the influence of annuli ring parameters, we modelled four ULE annular rings attached to the four mirrors. The annuli ring of outer radius the same as the mirror radius 6.35 mm, inner radius 3 mm and thickness of 4 mm unless they are varied. Figure \ref{fig8} (a) shows that as the thickness of the annular ring varied from 1 mm to 6 mm, a saturation behaviour was observed in both the first zero-crossing cutouts and the sensitivity values. From 4mm thickness above, no further change in zero-crossing or sensitivity.\\

As the annuli ring inner radius is varied from 1 to 6 mm, the behaviour is as shown in the Figure \ref{fig8} (b). The rate of change of zero-crossing and sensitivity increases and decreases.

\section{Impact of machining errors on cavity length stability}
\label{sec4}

Given the clear dependence of cavity length stability on geometric parameters, it is important to consider the implications of machining errors during the manufacturing process. Any machining errors in these parameters can directly affect the length stability of the cubic cavity. In this section, we estimate the resulting length instability caused by such errors, focusing on two critical parameters that significantly influence stability: the cutout depth and the bore radius. This will help us assess the potential impact of manufacturing tolerances on the overall performance of the ultra-stable cavity.\\

The extent of length instability introduced by machining errors depends on several factors, including the magnitude of the support forces ($F$) and the amount of dimensional error (which is determined by the machining tolerance). \\

The support forces required to securely mount an optical cavity strongly depend on the intended application. For space-based systems \cite{Cole:24} (e.g., optical atomic clocks on satellites, space interferometry, laser communication, radar payloads, or fundamental physics missions), launch and operational environments may impose accelerations of $10-50$ $g$, requiring support forces in the range of 250 N or above, depending on cavity mass. For terrestrial or mobile applications (e.g., transportable clocks, geophysical sensing, radar, or field metrology), accelerations are typically limited to $0.1-1$ $g$ and occasionally up to $\sim3$ $g$, necessitating secure support in the $20-100$ N range, again dependent on cavity size and mass.\\

Machining tolerance specifies the maximum permissible deviation from the intended dimension. The amount of tolerance achieved is highly dependent on the type of machining technology used. For ultra-low expansion (ULE) materials, one of the most suitable methods, like Selective Laser Etching (SLE) \cite{casamenti2024high}, can routinely achieve sub-micron to micron-level tolerances, typically around 0.001 mm. By contrast, more conventional machining methods, such as standard CNC milling or turning, usually achieve only $\sim 0.05-0.1$ mm precision. In our analysis, we therefore consider both the best-case (0.001 mm) and worst-case (0.1 mm) scenarios to estimate the maximum potential impact of machining tolerances on the dimensional stability of the cavity.

\subsection{\texorpdfstring{Errors in cutout depth ($\boldsymbol{\mathit{\Delta d}}$)}{Errors in Cutout depth (Δd)}}

To examine the influence of machining errors in the cutout depths on the stability of the cavity length, we consider the case where one of the four supporting cutouts has a machining error of $\Delta d=+0.001$ mm (for best-case tolerance) and $\Delta d=+0.1$ mm (for worst-case tolerance). We then analyze how this deviation impacts the length stability under varying levels of supporting force. The corresponding results are summarized in the Table \ref{tab4} below.\\

Our findings reveal that increasing the supporting force leads to a greater length instability to cutout depth errors. Importantly, we observe nearly 1-2 orders of improvement in the instability caused by best-case and worst-case machining tolerances irrespective of the applied supporting force.

\begin{table}[!h]
\centering
\renewcommand{\arraystretch}{1.3}  % increase row height for better spacing
\setlength{\tabcolsep}{5pt}        % horizontal padding inside cells

\begin{tabular}{|c|c|c|}
\hline
\begin{tabular}[c]{c}
\textbf{Supporting} \\ \textbf{force ($\boldsymbol{F}$)}
\end{tabular} &
\begin{tabular}[c]{c}
\textbf{Maximum} \\ \textbf{instability} \\ \textbf{(Best-case$^{a}$)}
\end{tabular} &
\begin{tabular}[c]{c}
\textbf{Maximum} \\ \textbf{instability} \\ \textbf{(Worst-case$^{b}$)}
\end{tabular} \\
\hline

1 N   & $1.922\times10^{-14}$ & $4.6\times10^{-13}$ \\
5 N   & $9.611\times10^{-14}$ & $2.3\times10^{-12}$ \\
10 N  & $1.922\times10^{-13}$ & $4.6\times10^{-12}$ \\
25 N  & $4.805\times10^{-13}$ & $1.2\times10^{-11}$ \\
50 N  & $4.723\times10^{-13}$ & $2.6\times10^{-11}$ \\
100 N & $8.683\times10^{-13}$ & $5.2\times10^{-11}$ \\
250 N & $2.171\times10^{-12}$ & $1.3\times10^{-10}$ \\
500 N & $4.342\times10^{-12}$ & $2.6\times10^{-10}$ \\
\hline
\end{tabular}

\caption{Instability analysis of the error in one of the cutout depths ($\Delta d$) under various supporting forces ($F$).}

\vspace{2mm}

{\footnotesize
$^{a}$ The best-case tolerance scenario considers a cutout depth of 12.969 mm with a $\Delta d = +0.001$ mm error.\\
$^{b}$ The worst-case tolerance scenario considers a cutout depth of 13.0 mm with a $\Delta d = +0.1$ mm error.
}
\label{tab4}
\end{table}

\subsection{\texorpdfstring{Errors in bore radius ($\boldsymbol{\mathit{\Delta b}}$)}{Errors in bore radius (Δb)}}

To examine the impact of machining errors in bore radius on the cavity length stability, we modelled the cubic cavity with one of the three bores that had a machining error $\Delta b$ in the bore radius and compute the length instability caused due to this deviation.\\

The simulation results presented in Table \ref{tab5} show that cavities subjected to larger supporting forces exhibit greater length instability. We observe that the best-case tolerance of 0.001 mm yields nearly an order of magnitude improvement in length stability compared to the worst-case tolerance of 0.1 mm, irrespective of the supporting forces.\\

\begin{table}[!h]
\centering
\renewcommand{\arraystretch}{1.3}
% Increase horizontal padding inside cells
\setlength{\tabcolsep}{5pt} 
\begin{tabular}{|c|c|c|}
\hline
\begin{tabular}[c]{c}
\textbf{Supporting} \\ \textbf{force ($\boldsymbol{F}$)}
\end{tabular} &
\begin{tabular}[c]{c}
\textbf{Maximum} \\ \textbf{instability} \\ \textbf{(Best-case$^{a}$)}
\end{tabular} &
\begin{tabular}[c]{c}
\textbf{Maximum} \\ \textbf{instability} \\ \textbf{(Worst-case$^{b}$)}
\end{tabular} \\
\hline

1 N   & $1.217\times10^{-13}$ & $1.4\times10^{-12}$ \\
5 N   & $6.083\times10^{-13}$ & $7.0\times10^{-12}$ \\
10 N  & $1.217\times10^{-12}$ & $1.4\times10^{-11}$ \\
25 N  & $3.042\times10^{-12}$ & $3.5\times10^{-11}$ \\
50 N  & $6.083\times10^{-12}$ & $7.0\times10^{-11}$ \\
100 N & $1.217\times10^{-11}$ & $1.4\times10^{-10}$ \\
250 N & $3.042\times10^{-11}$ & $3.5\times10^{-10}$ \\
500 N & $6.083\times10^{-11}$ & $7.0\times10^{-10}$ \\
\hline
\end{tabular}
\caption{Instability analysis of the error in one of the bore radii ($\Delta b$) under various supporting forces ($F$).}
\vspace{2mm}
{\footnotesize
$^{a}$ The best-case tolerance scenario considers a bore radius of 2.550 mm with a $\Delta b = +0.001$ mm error.\\
$^{b}$ The worst-case tolerance scenario uses a 2.5 mm bore radius with a $\Delta b = +0.1$ mm error.
}
\label{tab5}
\end{table}

\begin{figure}[htb]
 \centering
 \includegraphics[width=0.8\linewidth]{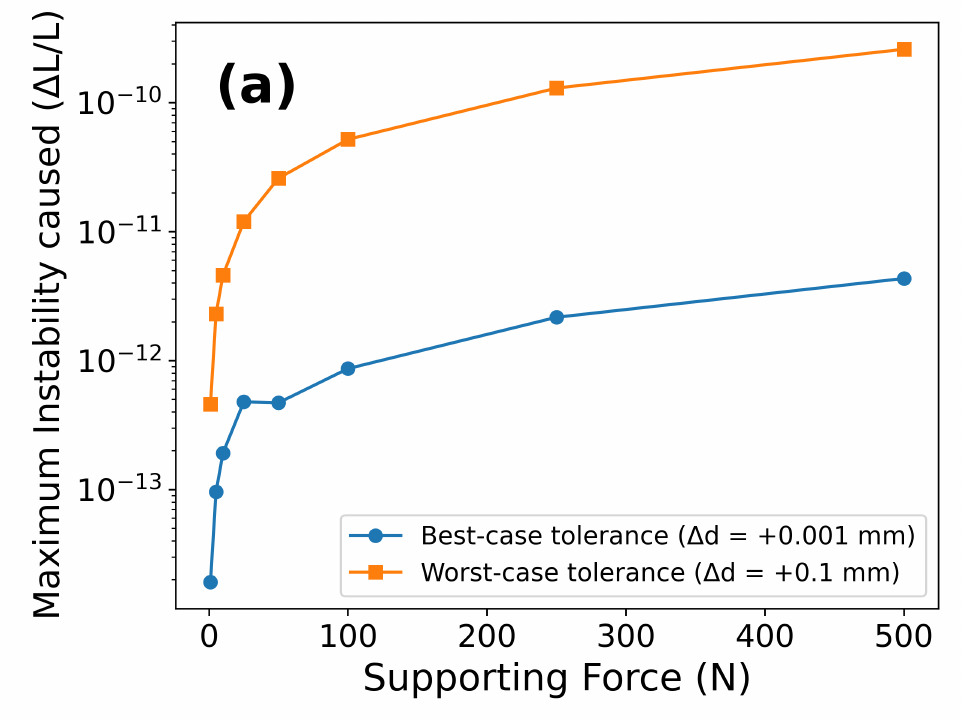}
 \includegraphics[width=0.8\linewidth]{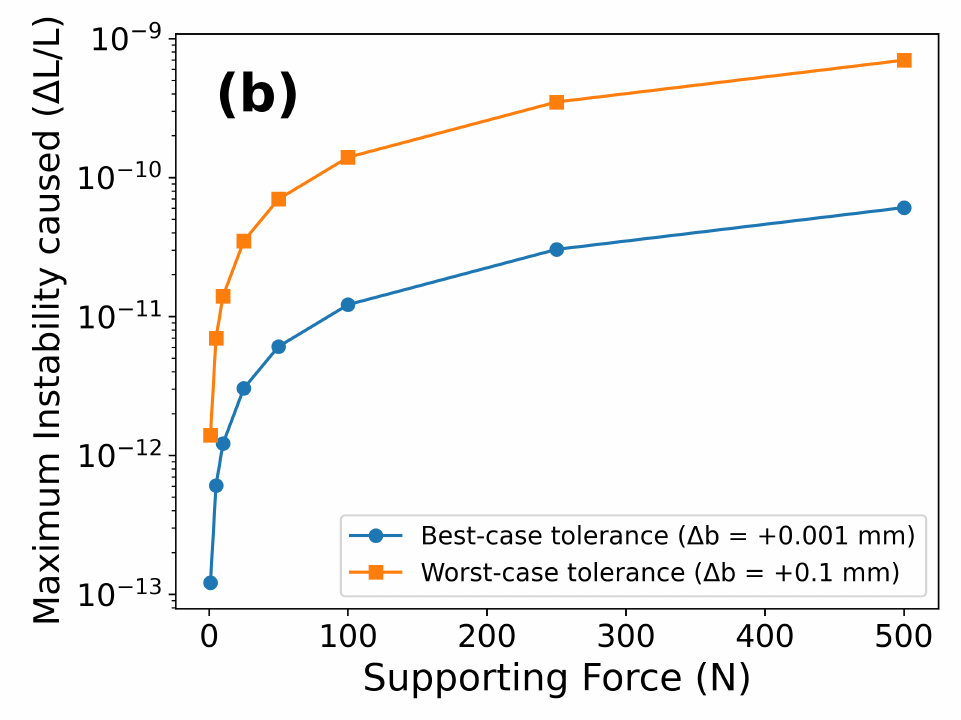}
 \caption{Effects of machining errors: (a) in cutout depth $\Delta d$ and (b) in bore radius $\Delta b$, plotted from the instability analysis in Table \ref{tab4} and Table \ref{tab5}. The lines in these figures are provided for visual guidance only and do not represent relationship between the data points.}
 \label{del_d&del_b}
\end{figure}

Both sets of results from Table \ref{tab4} and Table \ref{tab5} are plotted in Figure \ref{del_d&del_b}. These results emphasize the critical importance of achieving micron-level machining precision to significantly minimize cavity length instability.

\section{Thermal simulations} \label{sec5}

Fluctuations in environmental temperature significantly affect the stability of the optical path length in an ultra-stable cavity system by causing changes in the dimensions of the cavity spacer and mirrors. While vibrational noise seems to affect mostly at short time scales, temperature fluctuation noise mainly induces low-frequency noise, leading to long-term non-linear frequency drift of the stabilized laser \cite{pal2024transportable}. To minimize this, cavities are usually made of such materials that possess a zero crossing coefficient of thermal expansion (CTE) at a specific temperature. One common choice for room temperature cavity spacer is Ultra-low expansion (ULE) glass, which exhibits a CTE on the order of 10$^{-10}$/K around zero crossing temperature. To ensure that the cavity system operates near its zero-crossing temperature, meticulous thermal control measures are implemented.\\

A combination of active and passive thermal management strategies is used to establish and maintain the thermal equilibrium near the zero-crossing temperature. Active stabilization, achieved through Peltier elements, enables precise control of the cavity temperature. Complementing this, passive thermal shields reduce the impact of external temperature fluctuations. This multi-layered configuration ensures that the passive system attenuates rapid temperature changes, allowing the active stabilization mechanism sufficient time to respond effectively to the slower thermal variations.\\

In this section, we examine how the careful selection and design of passive thermal shielding can significantly increase the system’s thermal time constant, thereby improving cavity length stability against external thermal perturbations. Since the cavity operates within a vacuum chamber, convective heat transfer is negligible. Therefore, our analysis focuses on heat transfer through thermal conduction and radiation. Additionally, due to the high finesse of the optical cavity (Finesse $>$ 100,000), mirror heating from laser absorption becomes an important factor influencing the overall thermal dynamics. Our numerical thermal modelling and simulation using COMSOL Multiphysics incorporates thermal conduction, radiative heat transfer, and localized heating from the mirrors to predict the temperature evolution across different components and to evaluate the resulting thermal time constant of the system.

\subsection{Thermal time constant simulation}

The thermal stability of the reference cavity plays a crucial role in determining the achievable laser linewidth and long-term frequency drift. Therefore, accurate modelling of the thermal response of the cavity and the housing design becomes essential to predict its performance under environmental perturbations and to design effective thermal shielding strategies. In the present work, we initially evaluate the conduction-limited (since the cavity system is operating under vacuum, heat transfer via convection is almost negligible) thermal time constant of an ultra-stable cavity through a numerical simulation \cite{hagemann-ultrastable, dai2015thermal}. The model describes heat transfer in a cubic cavity enclosed within three highly polished thermal shields and a vacuum chamber, with conduction occurring through mechanical supports. This approach enables estimation of the system’s intrinsic thermal filtering capability in the absence of radiative effects and mirror heating. Subsequently, mirror heating was incorporated as a localized heat source in the numerical simulation. Finally, the combined effects of heat conduction, radiation, and mirror heating were evaluated based on the FEA approach using COMSOL to determine the effective thermal time constant.\\

\begin{figure*}[htb]
 \centerline{\includegraphics[width=0.35\linewidth]{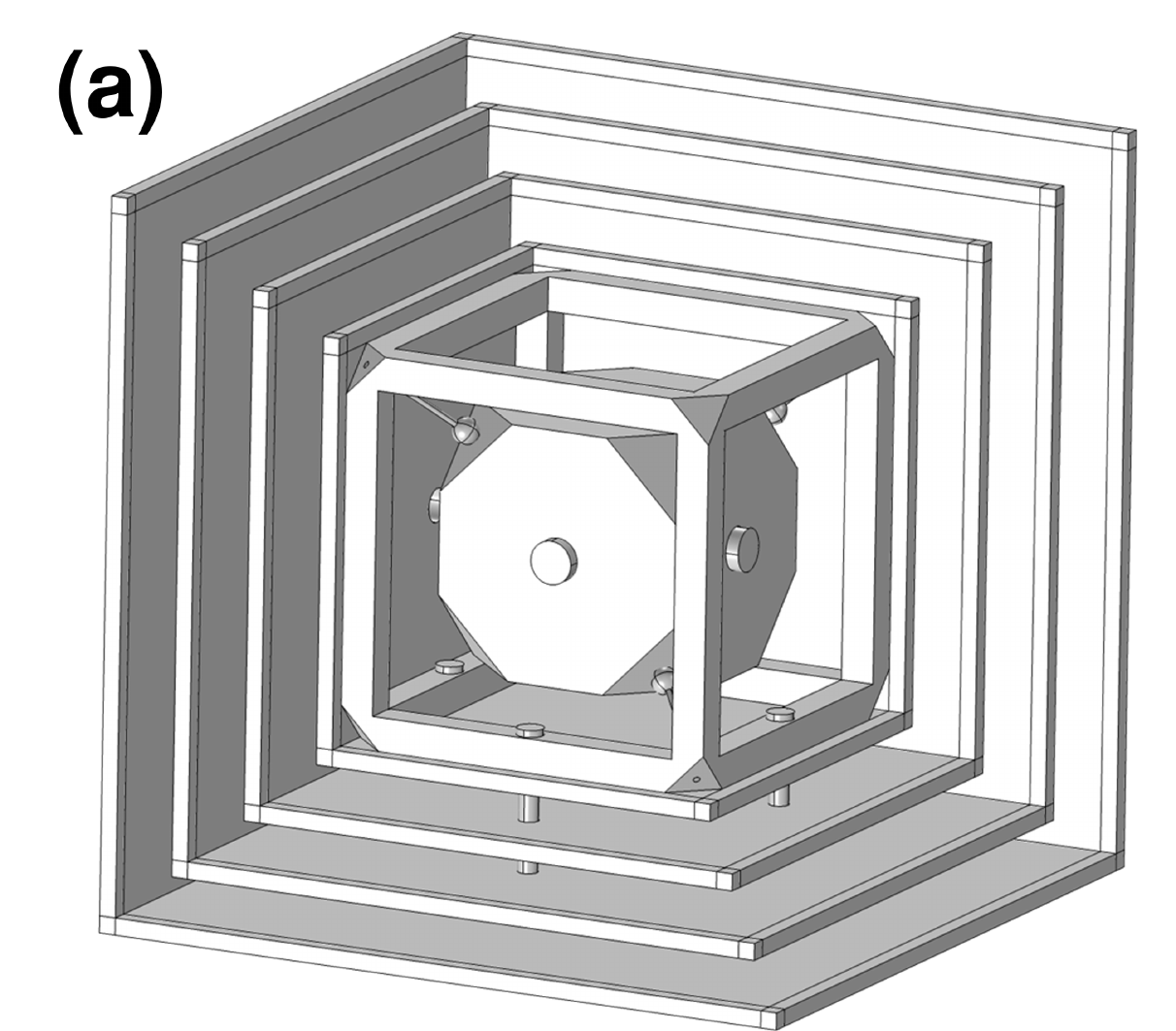}
 \includegraphics[width=0.55\linewidth]{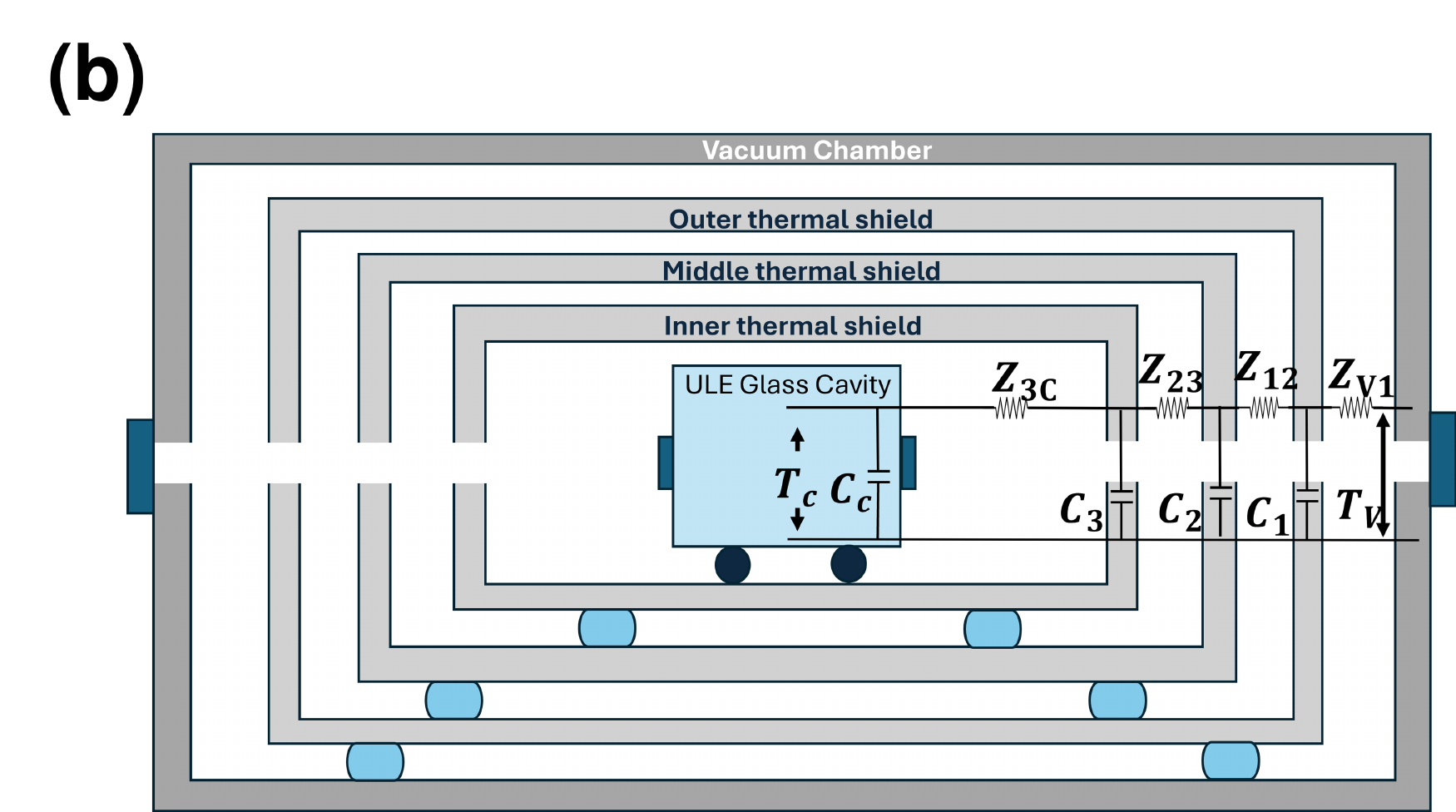}}
 \caption{(a) FEA model of cavity enclosed within three thermal shields and a vacuum chamber. (b) The effective thermal model.}
 \label{fig:thermal_model}
\end{figure*}

In the numerical modelling, the thermal shields were represented as thermal capacitors, while the gaps between successive shields, mechanically supported by glass spheres, were modelled as thermal resistors. For simplicity, the glass spheres were approximated as cylindrical connectors with minimal contact area, since point contacts dominate the heat transfer between shields in this configuration.\\

Under these assumptions, the thermal impedance is expressed as 
\begin{equation}
 Z_{\mathrm{th}} = \frac{L}{K A},
\end{equation}
where, $L$ is the length of the cylinder, $A$ is the cross-sectional area, and $K$ is the thermal conductivity of the material.\\

The thermal capacitance of the shields, modeled as cubic structures, is given by 
\begin{equation}
 C_{\mathrm{th}} = m C_p = v \rho C_v,
\end{equation}
where $m$ is the mass, $v$ is the volume, $\rho$ is the density, and $C_p$ and $C_v$ are the specific heats at constant pressure and constant volume, respectively.\\

To determine the time evolution of the temperature in different components of the ultra-stable cavity system, we applied the thermal-electrical analogy, treating the heat flow from the environment to the inner shields and cavity as a cascade of low-pass filters (Fig. \ref{fig:thermal_model} (b)). The governing equations below for this model yield the temporal temperature profiles for each component, enabling quantitative estimation of the conduction-limited thermal time constant \cite{hagemann-ultrastable, dai2015thermal}. In both numerical and FEA based method the whole system is initially taken as 293.15 K and then the outer side temperature of the vacuum chamber is raised by 1 K to find out the effective time constant of heat transfer.

\begin{equation}
\frac{d T_{s 1}}{d t}=\frac{T_{v}-T_{s 1}}{Z_{v 1} C_{1}}+\frac{T_{s 2}-T_{s 1}}{Z_{12} C_{1}} 
\end{equation}

\begin{equation}
\frac{d T_{s 2}}{d t}=\frac{T_{s 1}-T_{s 2}}{Z_{12} C_{2}}+\frac{T_{s 3}-T_{s 2}}{Z_{23} C_{2}} 
\end{equation}

\begin{equation}
\frac{d T_{s 3}}{d t}=\frac{T_{s 2}-T_{s 3}}{Z_{v 1} C_{3}}+\frac{T_{c}-T_{s 3}}{Z_{3 c} C_{3}} 
\end{equation}

\begin{equation}
 \frac{d T_{c}}{d t}=\frac{T_{s 3}-T_{c}}{Z_{3 c} C_{c}}
 \label{eq6}
\end{equation}

\begin{figure}[htb]
 \centering
 \includegraphics[width=0.9\linewidth]{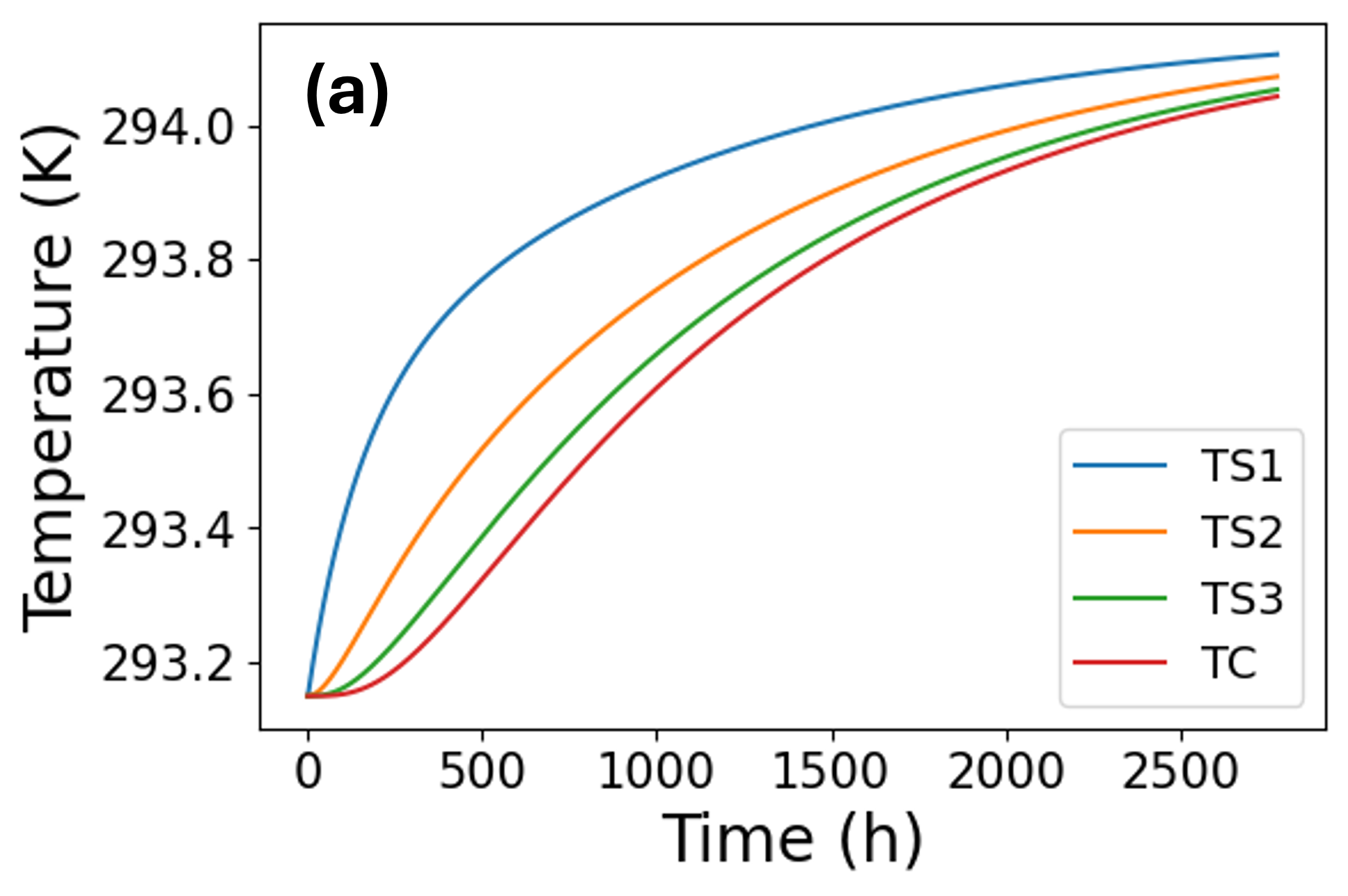}
 \includegraphics[width=0.9\linewidth]{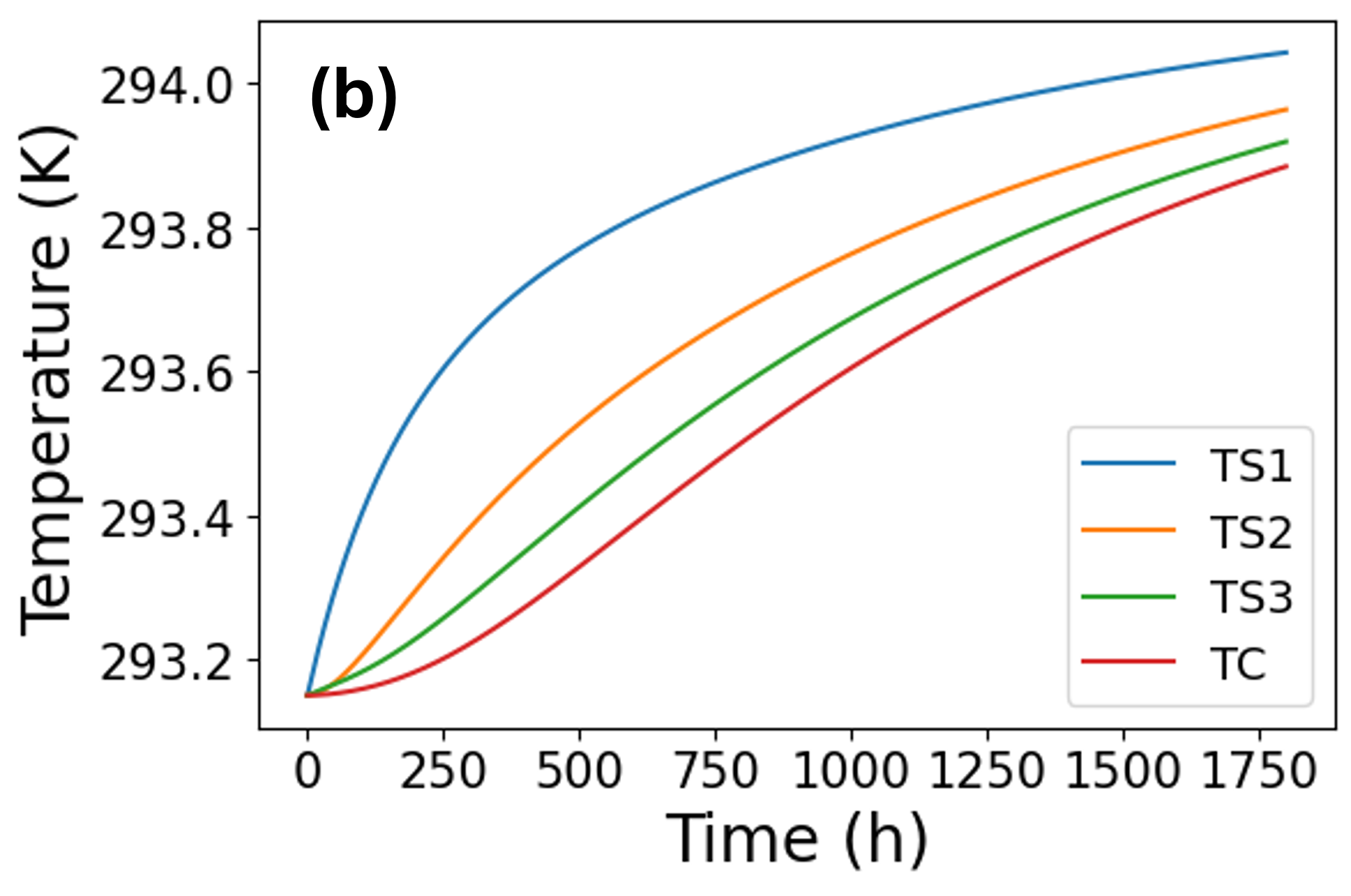}
 \caption{Time evolution of temperature of different parts of the cavity assembly when only heat conduction is taken into account (a) using numerical analysis method. (b) using FEM-based method}
 \label{fig11}
\end{figure}

Here, $\mathrm{T}_{\mathrm{v}}, \mathrm{T}_{\mathrm{S} 1}, \mathrm{~T}_{\mathrm{S} 2}, \mathrm{~T}_{\mathrm{S} 3}, \mathrm{~T}_{\mathrm{C}}$ are the temperatures of inner side of vacuum chamber, outer shield, middle shield, inner shield and cavity.
Table \ref{tab6} summarizes the thermal impedances and thermal capacitances associated with the individual components of the system. Figures \ref{fig11} (a) and \ref{fig11} (b) show the corresponding time evolution of temperature for different parts of the cavity assembly, obtained using numerical modelling and finite element method (FEM), respectively. In both approaches, the system is initially assumed to be in thermal equilibrium at 293.15K. A step increase of 1 K is then applied to the outer surface of the vacuum chamber in order to determine the effective thermal time constant, defined as the time required for the average temperature of the cavity components to reach $(1 - 1/e)\,\Delta T$.\\

When heat transfer is considered solely through conduction, the thermal time constants obtained from numerical modelling are approximately 1317.6 hours ($\approx$ 54.9 days) for the inner thermal shield and 1426.9 hours ($\approx$ 59.5 days) for the cavity spacer. Using the FEM based approach, the corresponding thermal time constants are found to be 1289 hours ($\approx$ 53.7 days) for the inner thermal shield and 1442 hours ($\approx$ 60.1 days) for the cavity spacer, in good agreement with the numerical results. The small discrepancy between the two methods is likely attributable to geometric simplifications employed in the numerical model, whereas the FEA simulations incorporate the full and exact geometry of the system. These exceptionally long thermal time constants demonstrate the strong passive thermal filtering capability of the system under purely conductive heat transfer, highlighting its effectiveness in maintaining cavity temperature stability over extended timescales in the absence of additional heat exchange mechanisms.\\

% \begin{table}[!h]
% \label{tab7}
% \caption{Different thermal components and their value depending upon geometry and material properties. Separation between thermal shields is 5 mm and the radius of the supporting PEEK cylinders are 3.5 mm.}
%     \centering
%     \renewcommand{\arraystretch}{2}
%     \setlength{\arrayrulewidth}{1pt}
%         \begin{tabular}{|c|c|>{\raggedright\arraybackslash}p{6cm}|}
%             \hline
%         \textbf{Thermal Components} & \textbf{Value} & \textbf{Description} \\
%         % \midrule
%         $\mathrm{Z}_{\mathrm{V}1}$ & $433.1\,\mathrm{W/K}$ & Resistance between outer thermal shield and vacuum shield \\
%         $\mathrm{Z}_{12}$ & $433.1\,\mathrm{W/K}$ & Resistance between outer thermal shield and second thermal shield \\
%         $\mathrm{Z}_{23}$ & $433.1\,\mathrm{W/K}$ & Resistance between middle thermal shield and inner thermal shield \\
%         $\mathrm{Z}_{3\mathrm{C}}$ & $547.13\,\mathrm{W/K}$ & Resistance between inner thermal shield and cavity \\
%         $\mathrm{C}_{1}$ & $2358.2\,\mathrm{J/K}$ & Thermal capacitance of outer thermal shield \\
%         $\mathrm{C}_{2}$ & $1636.58\,\mathrm{J/K}$ & Thermal capacitance of middle thermal shield \\
%         $\mathrm{C}_{3}$ & $1046.1\,\mathrm{J/K}$ & Thermal capacitance of inner thermal shield \\
%         $\mathrm{C}_{\mathrm{c}}$ & $685.456\,\mathrm{J/K}$ & Thermal capacitance of cavity \\
%         % \bottomrule
%     \end{tabular}
% \end{table}

\begin{table*}[!t]
\centering
\renewcommand{\arraystretch}{1.3}  % adjust row height for spacing
\setlength{\tabcolsep}{10pt}        % horizontal padding

\begin{tabular}{|c|c|c|}
\hline
\textbf{Thermal Components} & \textbf{Value} & \textbf{Description} \\
\hline
$\mathrm{Z}_{\mathrm{V}1}$ & $433.1\,\mathrm{W/K}$ & Resistance between outer thermal shield and vacuum shield \\
$\mathrm{Z}_{12}$ & $433.1\,\mathrm{W/K}$ & Resistance between outer thermal shield and second thermal shield \\
$\mathrm{Z}_{23}$ & $433.1\,\mathrm{W/K}$ & Resistance between middle thermal shield and inner thermal shield \\
$\mathrm{Z}_{3\mathrm{C}}$ & $547.13\,\mathrm{W/K}$ & Resistance between inner thermal shield and cavity \\
$\mathrm{C}_{1}$ & $2358.2\,\mathrm{J/K}$ & Thermal capacitance of outer thermal shield \\
$\mathrm{C}_{2}$ & $1636.58\,\mathrm{J/K}$ & Thermal capacitance of middle thermal shield \\
$\mathrm{C}_{3}$ & $1046.1\,\mathrm{J/K}$ & Thermal capacitance of inner thermal shield \\
$\mathrm{C}_{\mathrm{c}}$ & $685.456\,\mathrm{J/K}$ & Thermal capacitance of cavity \\
\hline
\end{tabular}
\caption{Different thermal components and their value depending upon geometry and material properties. Separation between thermal shields is 5 mm, and the radius of the supporting PEEK cylinders is 3.5 mm.}

\label{tab6}
\end{table*}

% \begin{table}[t]
% \caption{Different thermal components and their values depending upon geometry and material properties. Separation between thermal shields is 5~mm and the radius of the supporting PEEK cylinders is 3.5~mm.}
% \label{tab7}
% \centering

% \begin{tabular}{c c p{3.2cm}}
% \hline\hline
% Thermal Component & Value & Description \\
% \hline
% $Z_{\mathrm{V}1}$ & $433.1\,\mathrm{W/K}$ & Resistance between outer thermal shield and vacuum shield \\
% $Z_{12}$ & $433.1\,\mathrm{W/K}$ & Resistance between outer thermal shield and second thermal shield \\
% $Z_{23}$ & $433.1\,\mathrm{W/K}$ & Resistance between middle thermal shield and inner thermal shield \\
% $Z_{3\mathrm{C}}$ & $547.13\,\mathrm{W/K}$ & Resistance between inner thermal shield and cavity \\
% $C_{1}$ & $2358.2\,\mathrm{J/K}$ & Thermal capacitance of outer thermal shield \\
% $C_{2}$ & $1636.58\,\mathrm{J/K}$ & Thermal capacitance of middle thermal shield \\
% $C_{3}$ & $1046.1\,\mathrm{J/K}$ & Thermal capacitance of inner thermal shield \\
% $C_{\mathrm{c}}$ & $685.456\,\mathrm{J/K}$ & Thermal capacitance of cavity \\
% \hline\hline
% \end{tabular}
% \end{table}

% \begin{tablenotes}
% \item[$^{\rm a}$] The best-case tolerance scenario considers a bore radius of 2.550 mm with a $\Delta b=+0.001$ mm error.
% \item[$^{\rm b}$] The worst-case tolerance scenario uses a 2.5 mm bore radius with a $\Delta b=+0.1$ mm error.
% \end{tablenotes}

Since the laser is stabilized to a high-finesse optical cavity 
(\( \mathcal{F} > 10^{5} \)), the circulating intra-cavity intensity becomes significantly enhanced, following the relation \( I_{\text{inside}} \propto \mathcal{F}\, I_{\text{input}} \). This enhanced field intensity leads to localized heating of the cavity mirrors due to residual absorption. To account for this effect, Eq.~\eqref{eq6} must be modified, whereas the remaining equations governing the time evolution of different parts of the cavity assembly remain unchanged. The absorbed laser power at the mirror surfaces introduces a localized thermal load, which can be effectively modelled as a current source within the thermal equivalent circuit framework. This analogy is appropriate because, in the same way that a current source injects power into an electrical circuit, the absorbed optical power at mirror surface eventually injects heat into the whole assembly. Neglecting this contribution would underestimate both the effective thermal time constant and the long-term stability of the cavity, making its inclusion essential for an accurate description of system dynamics. Incorporating this representation yields the following modified form of Eq.~\eqref{eq6} \cite{J.yu2023cryogenic}:

\begin{equation}
 \frac{d T_{c}}{d t}=\frac{T_{s 3}-T_{c}}{Z_{3 c} C_{c}}+\frac{P_{\text{laser}}}{C_{c}}
\end{equation} 

\begin{figure}[htb]
 \centering
 \includegraphics[width=0.9\linewidth]{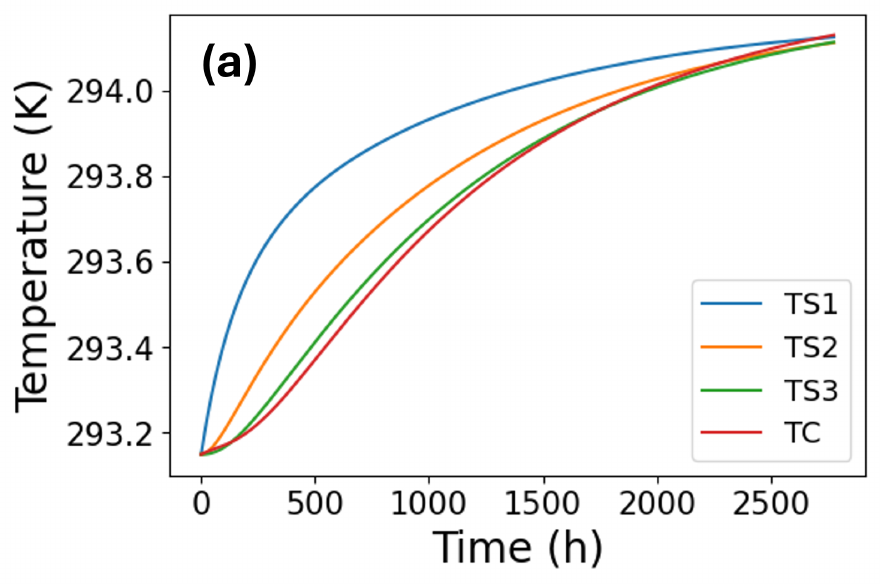}
 \includegraphics[width=0.9\linewidth]{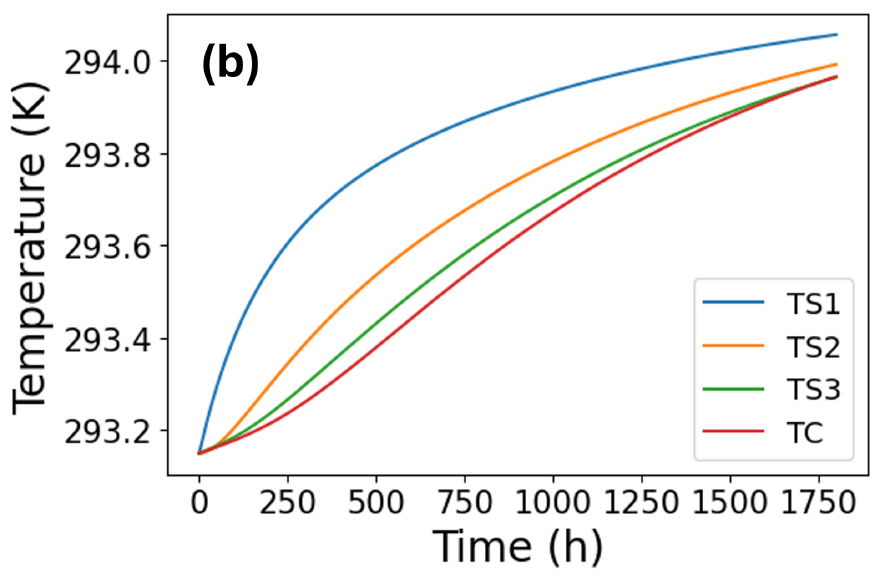}
 \caption{Time evolution of temperature of different parts of the cavity assembly when heat conduction and localized heating of mirror due to input laser beam are taken into account (a) using numerical analysis method. (b) using FEA based method}
 \label{fig12}
\end{figure}

Here, \(P_{\text{laser}}\) denotes the input laser power coupled into the cavity. When heat transfer through conduction and localized laser-induced heating are included in the numerical model, the resulting thermal time constants are approximately 1194.4 hours ($\approx$ 49.8 days) for the inner thermal shield and 1236.1 hours ($\approx$ 51.5 days) for the cavity spacer, for an input laser power of \(50~\mu\text{W}\) (Fig. \ref{fig12} (a)). Using the FEA-based method, the time constant of heat transfer for inner thermal shield and cavity are 1184 hours ($\approx$ 49.3 days) and 1242 hours ($\approx$ 51.8 days), which is in quite agreement with the numerical-based result. The possible reason for the small mismatch in the outcomes of numerical-based and FEA-based methods is given earlier. These results indicate that, under conduction and localized heating, the thermal shielding remains highly effective in suppressing external thermal perturbations.

Since the optical cavity operates under high vacuum conditions, a comprehensive thermal analysis was carried out by incorporating all relevant heat transfer mechanisms, namely conduction, radiation, and for the first time localized heating of mirrors, using finite element analysis to accurately estimate the effective thermal time constant. Upon inclusion of radiative heat transfer, the thermal time constant was found to reduce significantly to approximately 142 hours, corresponding to about six days. This result indicates that for room temperature ultra-stable optical cavity systems, radiative heat transfer constitutes the dominant thermal mechanism, as expected given that radiation governs heat exchange under high vacuum conditions of the order of $10^{-7}\,\mathrm{mbar}$. 
\begin{figure}[htb]
 \centering
 \includegraphics[width=0.9\linewidth]{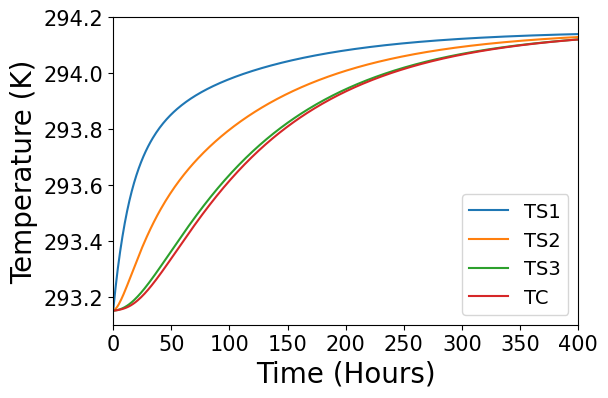}
 \caption{Time evolution of temperature of different parts of the cavity assembly when heat radiation, conduction and localized heating of the mirrors due to the input laser beam are taken into account}
 \label{fig13}
\end{figure}
Despite the reduced thermal time constant, the actively implemented thermal stabilization system based on a combination of Peltier coolers and resistive heaters successfully maintains the cavity at its zero crossing temperature. This demonstrates that the thermal shielding architecture is sufficiently robust to ensure stable operation even under stringent environmental conditions, thereby making the system suitable not only for controlled laboratory settings but also for outdoor and space based applications.

\subsection{Thermal expansion simulation}

In ultrastable optical cavities, temperature fluctuations induce thermal expansion in both the spacer and mirrors, directly affecting the length stability of the cavity. When fused silica mirror substrates are used, their relatively large coefficient of thermal expansion (CTE) compared to the ULE (ultra-low expansion) spacer material can cause significant deformation of mirrors. To mitigate this effect, Legero et al. \cite{Legero:10} proposed attaching ULE annular rings to the rear surface of the mirrors, thereby reducing the effective expansion mismatch and also allowing one to tune the zero-crossing temperature with annular ring parameters.\\

In this work, we carry out finite-element analysis (FEA) simulations to investigate the thermal expansion behaviour of a 7.5 cm dual-axis cubic cavity. In particular, we explore how the annular rings' parameters can influence the overall expansion response.\\

The fractional change in cavity length due to thermal expansion, for a temperature shift from $T_1$ to $T_2$, is expressed as:

\begin{equation}
 \frac{\Delta L}{L} = \int_{T_1}^{T_2} \alpha(T)\, dT
\end{equation}

where $\alpha(T)$ denotes the temperature-dependent coefficient of thermal expansion (CTE) of the material.

For the simulations, the CTEs of ULE and fused silica are expressed as:

\begin{equation}
 \alpha_{ULE}(T) = (T-T_0)(10^{-9}) - (T-T_0)^2(10^{-11}) 
\end{equation}
\begin{equation}
 \alpha_{FS}(T) = 0.55 \times 10^{-6} /K
\end{equation}
 where $T_0$ is the zero-crossing temperature of ULE, which is taken to be 20$\degree$C.\\

\begin{figure}[htb]
 \centering
 \includegraphics[width=0.8\linewidth]{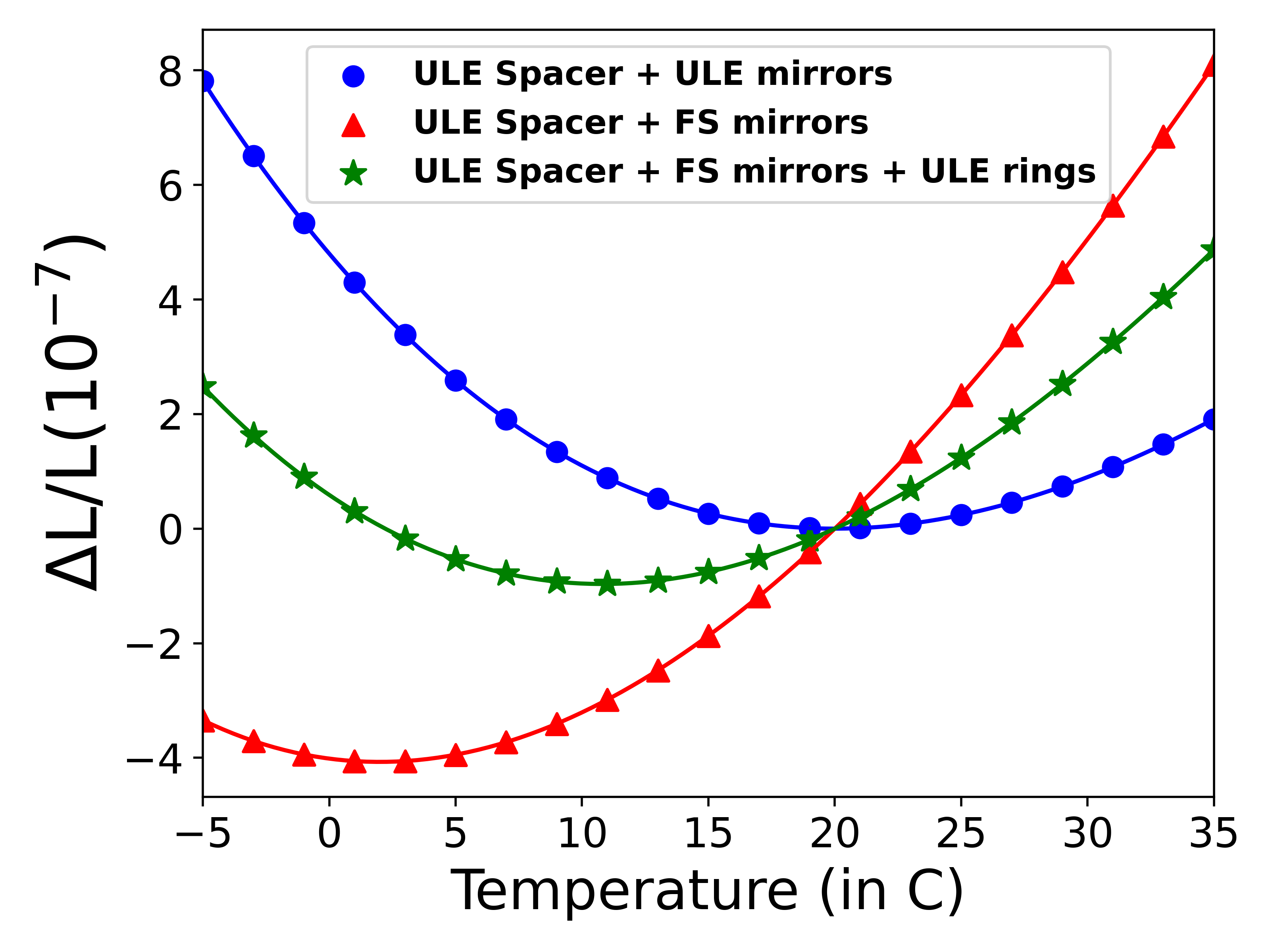}
 \caption{Thermal expansion behaviour simulated for three cases: a ULE spacer with ULE mirrors, a ULE spacer with fused silica (FS) mirrors, and FS mirrors equipped with ULE annuli rings.}
 \label{fig14}
\end{figure}

In the FEM model, the temperature variation is directly applied to the 7.5 cm dual-axis cubic cavity. The resulting length fluctuations are determined by probing the displacements of the mirrors caused by thermal expansion. The initial reference temperature is chosen as 20$\degree$C. The corresponding simulation results are plotted in Figure \ref{fig14}.\\

When ULE mirrors are replaced with FS mirrors, the minimum expansion temperature shifts from 20$\degree$C down to 1.86$\degree$C. But, by attaching ULE annuli rings to the FS mirrors, this temperature is increased again to 10.75$\degree$C. These results confirm the findings of Legero et al. \cite{Legero:10}, demonstrating that the use of ULE annular rings provides an effective compensation mechanism, shifting the minimum expansion temperature closer to room temperature.

\begin{figure}[htb]
     \centering
     \includegraphics[width=0.8\linewidth]{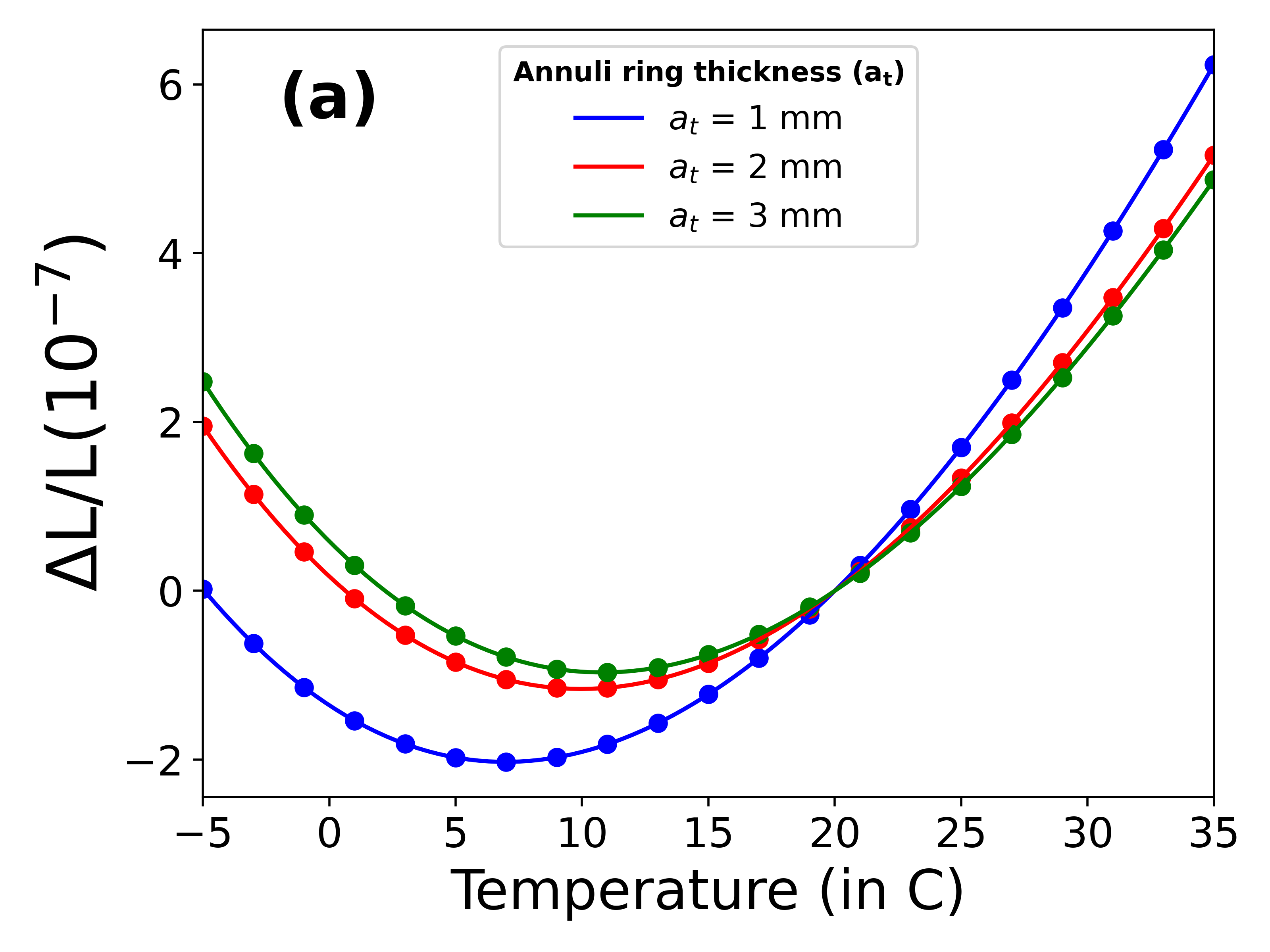}
     \includegraphics[width=0.8\linewidth]{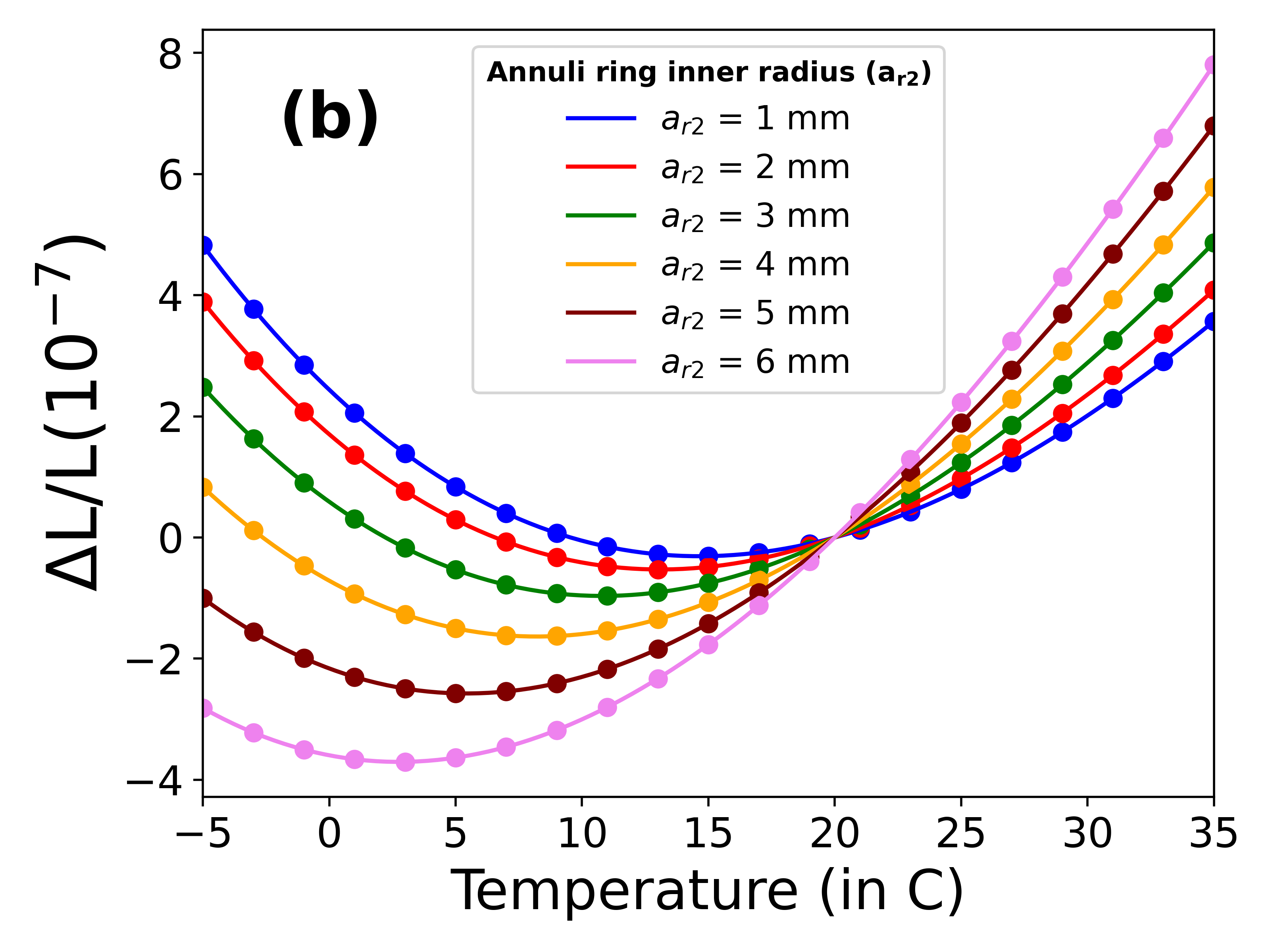}
     \caption{Thermal expansion behaviour as a function of (a) annuli ring thickness ($a_t$) and (b) annuli ring inner radius ($a_{r2}$).}
     \label{fig15}
\end{figure}

\subsubsection{Effect of annuli rings on thermal expansion}

Figure \ref{fig15} (a) examines the thermal expansion behaviour as the thickness of the annuli ring is varied from 1 mm to 3 mm. It is found that the minimum expansion temperature increases from 7.12$\degree$C to 9.94$\degree$C and then saturates at approximately 10.75$\degree$C as the annuli thickness becomes 3mm. This indicates that for an annuli thickness of 3 mm or greater, the minimum expansion temperature remains nearly constant and cannot be tuned further by increasing the thickness. Therefore, selecting a thickness of 3 mm is optimal and sufficient.\\

 Figure \ref{fig15} (b) shows the thermal expansion behaviour as the annuli ring inner radius is varied from 1 mm to 6 mm. The results show that the minimum expansion temperature decreases significantly with increasing inner radius of annuli rings. Specifically, the minimum expansion temperatures are 14.79$\degree$C, 13.18$\degree$C, 10.75$\degree$C, 8.33$\degree$C, 5.50$\degree$C, and 2.67$\degree$C for inner radii from 1 mm to 6 mm, respectively. A smaller annuli ring inner radius therefore shifts the minimum expansion temperature closer to room temperature, which is desirable for improved thermal stability. However, the inner radius cannot be made arbitrarily small, since sufficient clearance is required to allow the laser beam to pass between the mirrors. In practice, an inner radius greater than 3 mm is feasible and provides a balance between optical access and thermal stability. Overall, these results indicate that the inner radius of the annular rings has a stronger influence on length instability due to thermal expansion than the annuli thickness.

% These results demonstrate that by tuning the annular ring parameters(specifically thickness and inner radius), one can effectively control the zero-crossing temperature of the 7.5 cm cubic cavity, providing a practical means to compensate for the thermal expansion mismatch between fused silica mirrors and the ULE spacer.

\section{Conclusion}
In this work, we have presented a comprehensive finite-element analysis (FEA) of a 7.5 cm dual-axis cubic optical cavity, optimized for transportable optical clock applications. The cubic spacer design, supported tetrahedrally at truncated vertices, was shown to provide ultra-low acceleration sensitivity while maintaining mechanical robustness. Systematic investigations of the geometric parameters revealed that each parameter affects the optimal cutout depth and the overall fractional length stability differently, with the bore radius playing the most dominant role. Furthermore, the analysis of machining tolerances in cutout depth and bore radius highlights the importance of precision fabrication, especially under higher preload forces relevant for spaceborne or transportable scenarios.\\

Thermal simulations demonstrated that the multilayer shielded configuration, combined with vacuum housing, yields long thermal time constants on the order of several weeks, ensuring strong passive suppression of environmental perturbations. Our analysis shows that although radiative heat transfer remains the primary contributor to the thermal time constant of room-temperature cavities, the inclusion of laser-induced mirror heating in the model is essential. Accounting for such localized heat loads ensures a more accurate assessment of long-term stability, which is critical for the performance of high-finesse optical systems.\\

Overall, the 7.5 cm dual-axis cubic cavity represents a compact, mechanically robust, and thermally stable platform for frequency stabilization of multiple lasers. Its dual-axis configuration offers the added advantage of enabling simultaneous multi-wavelength operation, directly supporting advanced optical clock architectures. These results establish the proposed design as a strong candidate for next-generation transportable and space-qualified optical clocks, with potential applications spanning terrestrial PNT systems, geodesy, VLBI, and deep-space navigation.

\subsection*{Author Contributions}
Himanshu Miriyala and Rishabh Pal contributed equally to this work as co-first authors, finite element model development, finite element analysis, and manuscript writing. Arijit Sharma: conceptualization of the project, manuscript writing, review, editing, and overall project supervision.

\subsection*{Acknowledgments}

Rishabh Pal gratefully acknowledges financial support from the University Grants Commission (UGC, Govt. of India) through a Senior Research Fellowship (SRF). Himanshu Miriyala gratefully acknowledges the financial support provided by iNiF (IIT Tirupati Navavishkar I-Hub Foundation), IIT Tirupati. Arijit Sharma acknowledges funding received from IIT Tirupati Navavishkar I-Hub Foundation (iNiF) through the grant number IITTNiF/TDP/2024-25/P03. The authors especially thank Deepak Pandey from the Inter-University Centre for Astronomy and Astrophysics (IUCAA), Pune, for his critical reading and valuable feedback on the manuscript, which significantly enhanced the quality and clarity of this work. We also extend special thanks to Nadeem Ahmed, Sumit Achar, Vikrant Yadav and Prethivi T. for their assistance with the review of the manuscript.
\newline
\vspace{1cm}

\subsection*{Conflicts of Interest}
The authors declare no conflicts of interest.

\vspace{1cm}

\subsection*{Data Availability Statement}
The data that support the findings of this study are available from the corresponding author upon reasonable request.

\bibliography{ref}

\end{document}